%% file: main.tex
\documentclass[10pt,twocolumn,letterpaper]{article}

\usepackage{cvpr}              
\usepackage[accsupp]{axessibility}

\usepackage{booktabs}
\usepackage{multirow}
\usepackage{tabularx} 
\usepackage{array}
\usepackage{makecell}
\usepackage{graphicx}
\usepackage{amssymb}
\usepackage{xcolor}
\usepackage{caption}
\usepackage{tcolorbox}

\input{preamble}
\definecolor{cvprblue}{rgb}{0.21,0.49,0.74}
\usepackage[pagebackref,breaklinks,colorlinks,allcolors=cvprblue]{hyperref}

\def\paperID{*****} 
\def\confName{CVPR}
\def\confYear{2026}

\title{Unlocking Strong Supervision: A Data-Centric Study of General-Purpose Audio Pre-Training Methods}

\author{Xuanru Zhou\footnotemark[2]\\
Zhejiang University\\
Zhejiang, China\\
\and
Yiwen Shao\footnotemark[3]\\
Tencent AI Lab\\
Bellevue, USA\\
\and
Wei-Cheng Tseng\footnotemark[2]\\
UT-Austin\\
Austin, USA\\
\and
Dong Yu\\
Tencent AI Lab\\
Bellevue, USA\\
}

\begin{document}
\maketitle
\footnotetext[2]{Work done during an internship at Tencent AI Lab.}
\footnotetext[3]{Corresponding author: \texttt{yshao18@jhu.edu}.}

\input{sec/abstract}    
\input{sec/intro}
\input{sec/related_work}
\input{sec/method}

\input{sec/experiment}
\input{sec/conclusion}

\newpage
{
    \small
    \bibliographystyle{ieeenat_fullname}
    \bibliography{main}
}

\input{sec/suppl}


\end{document}

%% file: sec/abstract.tex
\begin{abstract}

Current audio pre-training seeks to learn unified representations for broad audio understanding tasks, but it remains fragmented and is fundamentally bottlenecked by its reliance on weak, noisy, and scale-limited labels. 
Drawing lessons from vision's foundational pre-training blueprint, we argue that the audio field must first establish its own large-scale, strong supervision framework.
We introduce a new data-centric pipeline that leverages a high-fidelity captioner to create SOTA-quality captions and the first Unified Tag System (UTS) that bridges speech, music, and environmental sounds. We then conduct a systematic comparative study of different pre-training objectives on these strong source data.
Our experiments suggest that data quality and coverage are the primary drivers of performance, while the choice of objective dictates downstream task specialization.

\end{abstract}

%% file: sec/intro.tex
\section{Introduction}
\label{sec:intro}

Large-scale pre-training has become the dominant paradigm for developing general-purpose models in computer vision field~\citep{imagenet-cls, CLIP, mehta2024catlipcliplevelvisualrecognition, rombach2021highresolution}. 
In comparison, the audio domain has historically been fragmented, with research siloed into specific sub-fields.
This has yielded powerful task-specific models for speech~\citep{wav2vec2, hsu2021hubert, whisper}, speaker-relevant~\citep{xvector, ecapa}, music~\citep{MERT, muq}, and general sounds~\citep{Panns, HTSAT,BEATs, ssast}.
While these models excel on respective benchmarks~\citep{superb, hear, marble}, this fragmentation has hindered the development of a single, unified model capable of spanning the full spectrum of audio data.

To achieve true general-purpose audio understanding~\citep{ALPsurvey}, Audio-Language Pre-training (ALP) has emerged as the most promising strategy. 
Inspired by the success of Vision-Language Models~\citep{CLIP,li2022blip, pmlr-v139-jia21b, siglip, siglip2}, ALP leverages free-form text as a universal supervisory signal. The prevalent approach is contrastive learning (CLAP)~\citep{CLAP2022, laionclap2023}, alongside other methods like generative captioning~\citep{tschannen2023cappa}, all aiming to learn representations that are both general-purpose and semantically rich.

However, this paradigm is fundamentally bottlenecked by its supervision source, which lacks both the scale of vision's multi-billion pair corpora~\citep{LAION5B, datacomp} and the quality of a true foundational dataset. Current audio caption datasets barely exceed one million pairs~\citep{audiosetcaps, wavcaps, audiocaps, clotho} and are often just ``LLM-augmented" sets, merely fluent rewrites of existing sparse tags~\citep{audiosetcaps, wavcaps}. This reliance on a noisy, semantically-sparse, and scale-limited source fundamentally caps pre-training potential. 
We argue that instead of mimicking the current web-scale VLM paradigm, the audio domain should first draw inspiration from an earlier, foundational blueprint from vision: the ImageNet paradigm~\citep{imagenet}. This approach proved that a large-scale, high-quality, strongly-supervised dataset could unify a field, and was later validated at massive scales via Multi-Tag Classification (MTC)~\citep{mtc, Mahajan_2018_ECCV}.
While PANNs~\citep{Panns} applied MTC in the audio domain, it was still limited by AudioSet~\citep{audioset}'s scope, which fails to truly unify the domains of speech, music, and environmental sounds.

\begin{figure*}[t]
    \centering
\includegraphics[width=\linewidth]{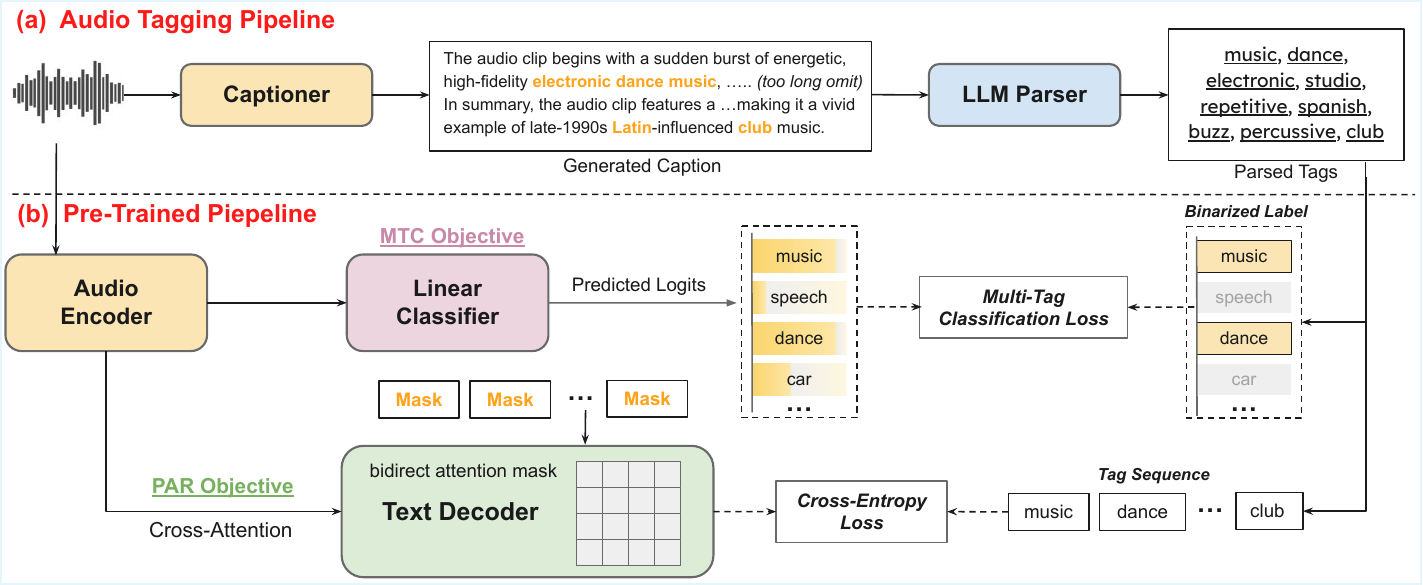}
    \caption{An overview of our method. (a) Audio Tagging Pipeline: we generate tags by processing raw audio through \textit{Qwen3-Omni-Captioner}~\citep{Qwen3-Omni} and an LLM Parser~\citep{qwen2025qwen25technicalreport}. (b) Pre-Training Pipeline: the audio encoder is then pre-trained using these tags with two supervision objectives: a discriminative multi-tag classification (MTC) Objective (using a linear classifier) and a generative parellel decoding (PAR) Objective (using a text decoder with a bidirectional mask).}
    \label{fig-pipeline}
\end{figure*}

To this end,  we introduce a new data-centric paradigm as shown in Fig.~\ref{fig-pipeline}. We leverage a high-fidelity audio captioner, \textit{Qwen3-Omni-Captioner}~\citep{Qwen3-Omni}, to process a diverse audio dataset \textit{CaptionsStew}~\citep{anonymous2025revisiting} 400k-subset.
From its rich, descriptive captions, we generate two distinct strong supervision sources: 
(1) a set of high-fidelity SOTA-quality captions for audio-language pre-training, and (2) a large-scale, data-driven Unified Tag System (UTS) for tag-oriented pre-training, which is the first of its kind, with up to 3k tags, that naturally bridges speech, music, and environmental sounds.
We then conduct a systematic comparative study of all pre-training objectives.
Our results are conclusive: (1) Our UTS-trained models achieve superior out-of-domain generalization (e.g., on speech/music) over the MTC (AudioSet) baseline, despite using 5x less data. (2) Our SOTA-Caps-trained models consistently outperform the~\citep{anonymous2025revisiting} baselines across a majority of tasks. This confirms our core hypothesis: the quality and richness of the supervision source is a primary driver of performance in audio pre-training.

Our contributions are threefold. First, we introduce a novel data-centric pipeline to create the first large-scale Unified Tag System (UTS) that bridges speech, music, and environmental sounds. 
Second, we conduct a systematic comparative study of different audio pre-training objectives, providing a comprehensive benchmark on this new strong supervision source. 
Finally, our results suggest that data quality and coverage can be more impactful than data volume, and that the choice of pre-training objective dictates downstream task specialization, offering a clear lesson for the future of audio pre-training.
To facilitate future research, we open-source the UTS data at \url{https://huggingface.co/datasets/AudenAI/UTS} and the accompanying code at \url{https://github.com/AudenAI/Auden/tree/main/examples/uts}.

%% file: sec/related_work.tex
\section{Related Work}
\label{sec:related-work}

\subsection{Audio Representation Learning}
Audio representation learning aims to develop a single, general-purpose model for diverse understanding tasks. This field has progressed along two parallel tracks: 
The first is task-specific supervised learning, which has produced high-performing encoders for audio event classification~\citep{VGGish, OpenL3, Panns, AST, HTSAT, Dasheng}, speech recognition systems~\citep{whisper}, and speaker recognition models~\citep{xvector, ecapa}. While these models are strong, their effectiveness is inherently limited by their specific, predefined label sets. 
The second major approach is self-supervised learning (SSL), which learns from unlabeled data and has yielded powerful, specialized encoders for speech~\citep{wav2vec2, hsu2021hubert, wavlm, data2vec}, general audio~\citep{ssast, audioMAE, BEATs, li2022atst}, and music~\citep{MERT, muq}. SSL models also show promise for generative tasks such as audio synthesis.
While these methods improve generalization within specific domains, the challenge remains to develop a representation that seamlessly integrates speech, music, and environmental sounds while overcoming task-specific limitations.

\subsection{Audio-Language Pre-Training}
Inspired by the success of Vision-Language Pre-training~\citep{CLIP,li2022blip, pmlr-v139-jia21b, siglip, siglip2}, Audio-Language Pre-training has emerged to learn rich, semantic features by leveraging the natural co-occurrence of audio and text.
The field is first dominated by Contrastive Learning~\citep{CLAP2022, CLAP2023, laionclap2023, guzhov2021audioclip}, with recent extensions have explored combinations with other objectives~\citep{xu2023blat, zhu2024cacophony, niizumi2024m2d, niizumi2025m2d2}, most notably Generative Pre-training (e.g., Captioning~\citep{tschannen2023cappa,midashenglm7b, af3}). 
This has also spurred an evolution in datasets. Influenced by the prevalence of using LLMs as caption re-writers in the vision field~\citep{clip-rewrite}, this trend has moved from small, human-annotated corpora~\citep{audiocaps, clotho, MusicCaps} to large-scale, ``LLM-augmented" collections~\citep{wavcaps, audiosetcaps, fusionaudio, autoacd} and domain-specific resources~\citep{paraspeechcaps, jamendomaxcaps}.
However, most existing datasets, even those ``LLM-augmented" are typically derived from existing sparse tags or noisy web-captions; the LLM is often used merely to rewrite these weak sources into fluent prose. Thus, the semantic richness of the supervision is still constrained by the low-quality original source.

\subsection{Vision's Blueprint for Supervision}
The methodologies from vision-based large-scale representation learning provide a clear blueprint for the audio domain. While recent Vision-Language Models (VLMs) have demonstrated the power of web-scale supervision, two key lessons from earlier vision work are particularly relevant.
First, the ImageNet~\citep{imagenet} paradigm highlighted the transformative effect of a single, large-scale, strongly-supervised dataset. For nearly a decade, pre-training on ImageNet’s 1000-class vocabulary was a dominant transfer learning strategy. More recent models, such as those in \citep{residual}, further established that a semantically-rich, classification-based dataset can be a powerful foundation for downstream tasks.
Second, Multi-Tag Classification (MTC)~\citep{mtc} proved to be an effective and scalable pre-training objective. \cite{Mahajan_2018_ECCV} demonstrated the power of training on billions of Instagram images and their noisy user-generated hashtags, showing that MTC can be a valuable source of strong supervision. 
Recent VLM~\citep{huang2024tag2text,liu2023tagalign, 10.1609/aaai.v37i1.25113, superclass_huang} show that MTC offers a more direct and structured semantic supervision signal and leads to stronger downstream performance in tasks such as image captioning and retrieval.
Our work builds on these vision strategies by applying them to the audio domain. Specifically, we aim to (1) establish a unified foundational audio classification dataset and (2) leverage MTC as the primary form of strong supervision for general audio.

%% file: sec/method.tex
\section{Method}
\label{method}

\subsection{Dataset Construction}
\label{sec:data-cons}
We utilize audio data from the \textit{CaptionStew} 400K-subset introduced in \cite{anonymous2025revisiting}, which encompasses a diverse range of source including \textbf{speech, music, and environmental sounds}. We then employ the \textit{Qwen3-Omni-Captioner}~\citep{Qwen3-Omni}, $f_{cap}$, to generate a detailed natural language caption $c_i = f_{\text{cap}}(a_i)$ for each audio clip $a_i$. These captions are then analyzed by a large language model (LLM), which extracts relevant tags based on the semantic content. This process results in a comprehensive and diverse tag vocabulary derived from the entire audio-caption dataset. Details of the data source can be found in Supplementary Material Sec.6.

\paragraph{LLM-Aided Tag Parsing.}
While conventional tag parsing methods in vision filed~\citep{Xu_2022_CVPR, xu2023learning, huang2024tag2text} often utilize the Natural Language Toolkit (NLTK)~\citep{bird-loper-2004-nltk}, this approach struggles with the verbosity of modern caption generators. The captions from \textit{Qwen3-Omni-Captioner}, averaging 388.43 words (see example in Supplementary Material Sec.7.1), are poorly served by NLTK's reliance on Part-of-Speech (POS) tagging, which hinders a full semantic comprehension.
To address this gap, we turn to Large Language Models (LLMs), which are inherently better suited for this challenge due to their advanced capacity for interpreting complex instructions. Inspired by~\cite{liu2023tagalign} that also employ LLMs for parsing image captions, we adopt \textit{Qwen2.5-7B-Instruct}~\citep{qwen2025qwen25technicalreport} as our tag parser, $g_{\text{parse}}$. This parser analyzes and distills a set of candidate tags $T'_i = g_{\text{parse}}(c_i)$ from each caption. The specific prompt used to instruct the LLM is detailed in Supplementary Material Sec. 7.2.
\vspace{-5pt}
\paragraph{Tag Category System Construction.}
\label{sec:uts}
The parsing process generates an audio-tag pair set $\{ (a_i, T'_i) \}_{i=1}^N$. We observed that the complete, aggregated tag vocabulary $\mathcal{T}_{\text{pool}} = \bigcup_{i=1}^N T'_i$ was too vast, creating a data sparsity problem that would hinder a multi-tag classification framework. 
Consequently, we chose to establish fixed-size tag systems, exploring different granularities where the vocabulary size $K \in \{800, 1\text{k}, 1.5\text{k}, 2\text{k}, 3\text{k}\}$.
The selection of these $K$ tags was guided by the TF-IDF algorithm~\citep{tf-idf}, as it provides a robust measure of a tag's importance by considering both its local frequency and its global distinctiveness. 
The score $s(t)$ for a tag $t \in \mathcal{T}_{\text{pool}}$ is:
\begin{equation}
\label{tf-idf}
 s(t) = df(t) \times \log \left( \frac{N+1}{df(t)+1} \right)   
\end{equation}
where $df(t)$ is the document frequency of tag $t$. We select the top-$K$ scoring tags to form our final Unified Tag System (UTS), $\mathcal{V} = \{v_1, \ldots, v_K\}$. This UTS is novel in that it unifies speech, music, and environmental sounds. This provides a crucial label foundation, making it possible to develop a single pre-training framework for encoders across all three audio modalities.
Further analysis of the unified tag system is available in Sec.~\ref{sec:dataset-analysis}.

\subsection{Tag-Oriented Pre-Training}

\paragraph{Motivation.}
Dominant audio-language objectives, such as CLAP~\citep{CLAP2022,CLAP2023, laionclap2023,learn-from-clip} or captioning~\citep{midashenglm7b, af3}, became prevalent because they cleverly use free-form text as a universal supervisory signal, bypassing the need for a unified label set across diverse data. However, their effectiveness has been limited by the quality of their data source; the free-text itself, often scraped from the human annotation~\citep{kim-NAACL-HLT-2019, clotho} or generated by LLMs~\citep{mei2023wavcaps}, is typically noisy, sparse, and provides only weak supervisory signals.
Our approach begins by solving this fundamental data problem. We derive our Unified Tag System (UTS) from captions generated by a high-fidelity audio captioner (as detailed in Sec.~\ref{sec:data-cons}). This process distills semantically rich, dense, and accurate information into a discrete tag set, creating a strong supervision source.
With this powerful UTS in hand, we can now apply inherently strong supervision objectives to pre-train our encoder. We explore two such methods: (1) Multi-Tag Classification (MTC)~\citep{mtc}, a direct, validated classification approach, and (2) Parallel Decoding~\citep{gu2018nonautoregressive}, a non-autoregressive generative objective. We now detail these two approaches.

\paragraph{Multi-Tag Classification Objective.}
\label{sec:mtc}
With the UTS $\mathcal{V}$ defined, we generate a $K$-dimensional, multi-hot binary label vector $y_i \in \{0, 1\}^K$ for each audio sample $a_i$. The $j$-th element of $y_i$ is defined as:
\begin{equation}
y_{ij} = \begin{cases} 1 & \text{if } v_j \in T'_i \\ 0 & \text{otherwise} \end{cases}    
\end{equation}
Samples where $y_i$ becomes a zero-vector are filtered out. This results in our final pre-training dataset $\mathcal{D} = \{(a_i, y_i)\}_{i=1}^{M}$, where $M \le N$.

We formulate the pre-training task as a standard multi-label classification problem. Our model, composed of an audio encoder $h_{\theta}$ and a linear classification head $W_c$, is trained to predict the probability $\hat{y}_i \in [0, 1]^K$ for all $K$ tags in our Unified Tag System. The entire model is trained end-to-end to minimize the Binary Cross-Entropy (BCE) loss $\mathcal{L}_{\text{MTC}}$. This objective compares the model's predictions $\hat{y}_i$ against the ground-truth multi-hot label vector $y_i$ for each sample $a_i$, and sum over all $K$ classes, averaged over all $M$ samples in the dataset $\mathcal{D}$:
\begin{equation}
\label{mtc}
    \mathcal{L}_{\text{MTC}} = - \frac{1}{M} \sum_{i=1}^{M} \sum_{j=1}^{K} \left[ y_{ij} \log(\hat{y}_{ij}) + (1 - y_{ij}) \log(1 - \hat{y}_{ij}) \right]
\end{equation}
After pre-training, the classification head $W_c$ is discarded. The learned encoder $h_{\theta^*}$ serves as our general-purpose audio representation model, which is then evaluated on a variety of downstream audio understanding tasks.

\paragraph{Generative Pre-training with Parallel Decoding.}
\label{sec:pd}
As an alternative to the MTC objective, we explore a generative pre-training approach that treats the tags as a structured text sequence. This method leverages an encoder-decoder architecture ($h_{\theta}$ and $g_{\phi}$) to explicitly model the mapping from audio to a textual representation of its tags.
First, for each audio clip $a_i$, we convert its ground-truth multi-hot vector $y_i$ back into a canonical text sequence $Y_i$. We construct this sequence by concatenating all positive tags from our Unified Tag System $\mathcal{V}$ using a comma separator (e.g., $Y_i = \texttt{"tag\_a, tag\_d, tag\_k"}$). This sequence $Y_i = (y_1, \ldots, y_T)$ becomes the target for a generative loss.

A standard autoregressive (AR) captioning objective~\citep{show-and-tell} (Eq.~\ref{eq:ar-loss}) conditions the prediction of the next token $y_t$ on both the audio $z_i^a = h_{\theta}(a_i)$ and the previously generated tokens $y_{<t}$. However, this autoregressive assumption is a poor fit for our task. The target $Y_i$ is a set of co-occurring tags where the order is arbitrary and carries no contextual logic or linguistic structure. Training an AR model on this data would force it to learn spurious sequential dependencies (e.g., that ``bark" must always follow ``dog"), which is both inefficient and incorrect.
\begin{equation}
\label{eq:ar-loss}
\mathcal{L}_{\text{ar}} = - \sum_{t=1}^{T} \log p_{\phi}(y_t \mid y_{<t}, z_i^a)
\end{equation}
To mitigate this, we adopt a Parallel Decoding objective, inspired by non-autoregressive models~\citep{gu2018nonautoregressive}. This approach is perfectly suited to our set-prediction task. In this mode, we mask the decoder's input sequence and remove the causal attention mask. This forces the model to predict all $T$ tokens simultaneously, conditioned solely on the audio features:
\begin{equation}
\label{eq:par-loss}
\mathcal{L}_{\text{par}} = - \sum_{t=1}^{T} \log p_{\phi}(y_t \mid z_i^a)
\end{equation}
This non-autoregressive objective compels the audio encoder $h_{\theta}$ to produce a much richer and more complete representation $z_i^a$, as this single vector must contain all information necessary to generate the entire tag sequence at once.
Our model is trained purely on this $\mathcal{L}_{\text{par}}$ objective, enforcing the strongest possible dependency on the audio encoder and forcing it to serve as the sole source of information for the tag generation task.

\subsection{Audio-Language Pre-Training}
As an orthogonal line of inquiry, we now investigate the value of the full semantic captions generated by \textit{Qwen3-Omni-Captioner}. Our hypothesis is that the high information quality of these new captions can directly boost the performance of standard audio-language objectives. We utilize the final summary paragraph of the generated caption as the text supervision $Y_i$, as the full-length captions are too long.

\paragraph{Contrastive Objective.}
Contrastive pre-training has proven to be a highly effective and scalable method for learning joint audio-language representations~\citep{CLAP2022, laionclap2023}. This objective trains two separate encoders, an audio encoder $h_{\theta}$ and a text encoder $h_{\psi}$, to project their respective outputs into a shared latent space. The encoders are optimized to maximize the similarity (e.g., cosine) between corresponding (audio, text) pairs ($z_i^a, z_i^t$) while simultaneously minimizing it for all other mismatched pairs in the batch. This is achieved using a symmetric InfoNCE loss~\citep{infonce}:
\begin{equation}
\label{eq:contrastive-loss}
\begin{split}
\mathcal{L}_{\text{con}} = -\frac{1}{2M}\sum_{i=1}^{M} \bigg[ & \log \frac{\exp(\text{sim}(z_i^a, z_i^t)/\tau)}{\sum_{j=1}^{M} \exp(\text{sim}(z_i^a, z_j^t)/\tau)} \\
& + \log \frac{\exp(\text{sim}(z_i^t, z_i^a)/\tau)}{\sum_{j=1}^{M} \exp(\text{sim}(z_i^t, z_j^a)/\tau)} \bigg]
\end{split}
\end{equation}

\paragraph{Captioning Objective.}
We also explore a generative pre-training objective. As discussed in Sec.~\ref{sec:pd}, autoregressive (AR) and parallel (PAR) decoding offer different trade-offs. Here, we adopt the mixed-training strategy from CapPA~\citep{tschannen2023cappa}. For each minibatch, a fraction of samples is trained with the standard autoregressive objective $\mathcal{L}_{\text{ar}}$ (Eq. \ref{eq:ar-loss}), while the remainder is trained with the non-autoregressive parallel objective $\mathcal{L}_{\text{par}}$ (Eq. \ref{eq:par-loss}). 
The full generative loss $\mathcal{L}_{\text{gen}}$ is the sum of these two components.

\paragraph{Multi-task Objective.}
We investigate whether the strong, discrete signal from our MTC objective can be beneficially combined with the dense, generative signal from captioning. We propose a multi-task learning (MTL) framework that jointly optimizes our two strongest objectives: the Multi-Tag Classification objective $\mathcal{L}_{\text{MTC}}$ (from Sec.~\ref{sec:mtc}) and the Generative Captioning objective $\mathcal{L}_{\text{gen}}$. The final loss is a weighted sum of the two, balanced by a hyperparameter $\lambda$:
\begin{equation}
\label{eq:mtl-loss}
\mathcal{L}_{\text{MTL}} = \mathcal{L}_{\text{MTC}} + \lambda\mathcal{L}_{\text{gen}}
\end{equation}
This MTL objective forces the audio encoder $h_{\theta}$ to learn representations that are simultaneously discriminative (for MTC) and descriptive (for captioning).

%% file: sec/experiment.tex
\section{Experiment}
\label{sec:exp}

\begin{figure*}[th]
  \centering
  \begin{subfigure}{0.3\linewidth}
    \includegraphics[width=\textwidth]{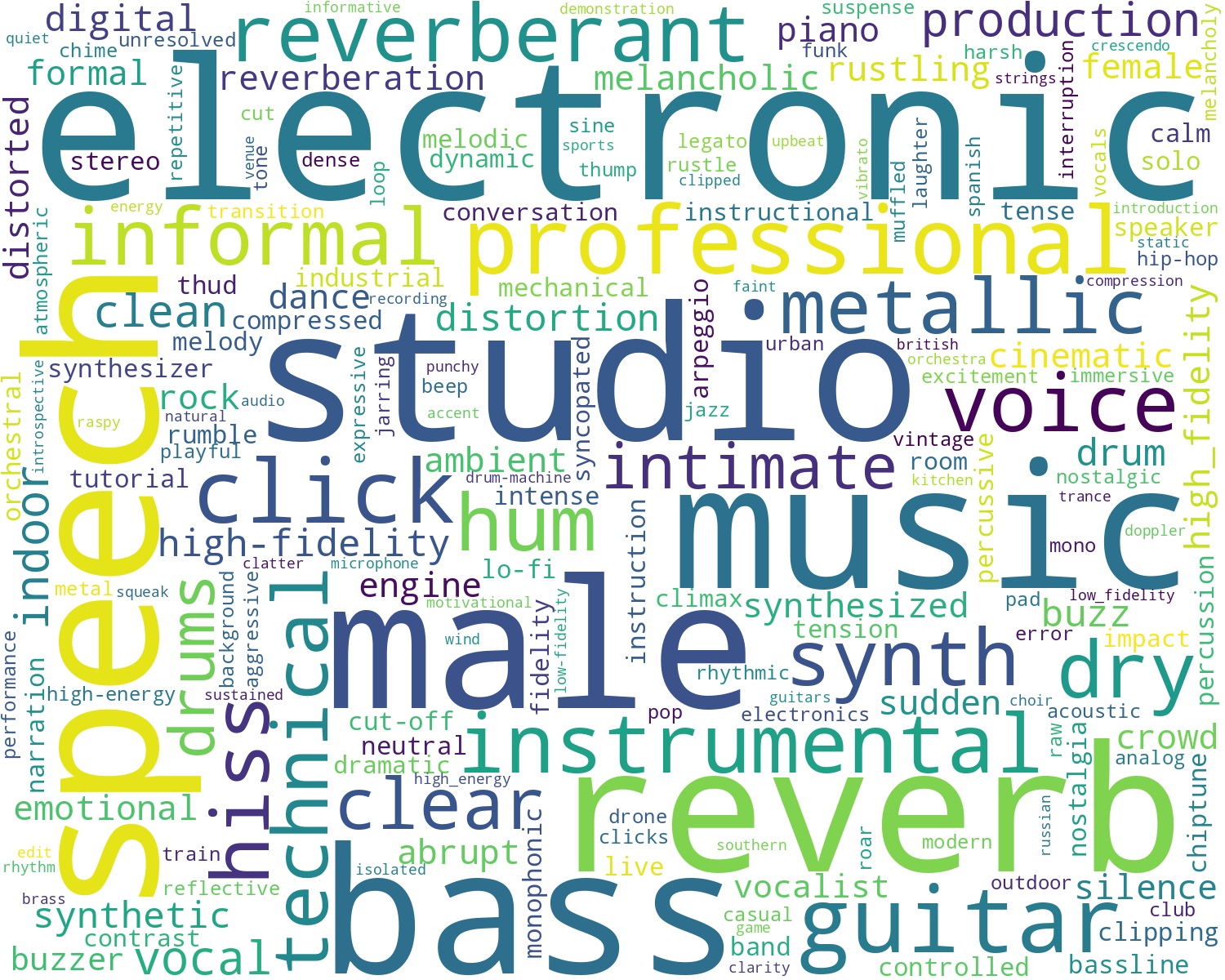}
    \caption{}
    \label{fig:wordcloud}
  \end{subfigure}
  \hfill
  \begin{subfigure}{0.32\linewidth}
    \includegraphics[width=\textwidth]{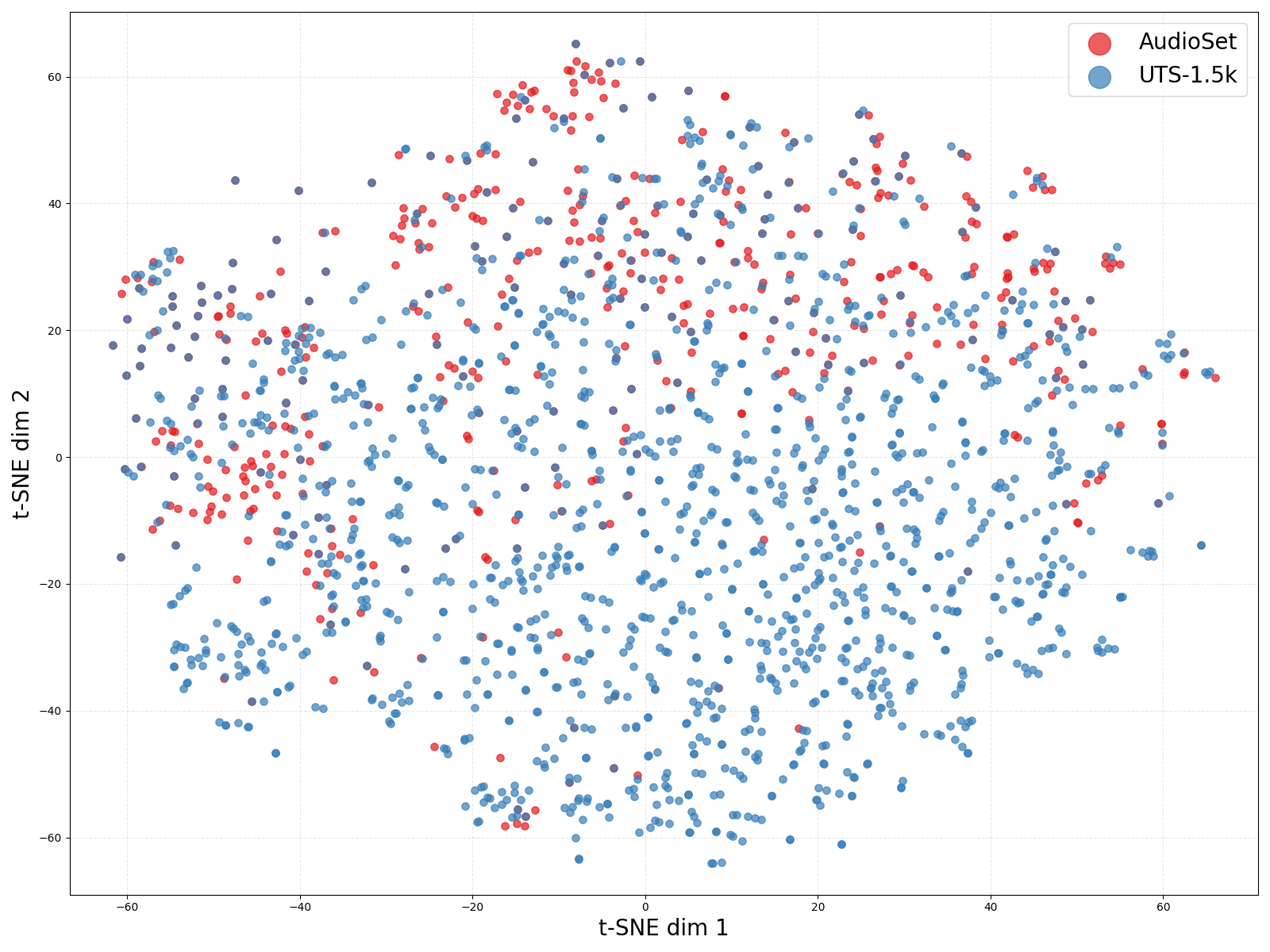}
    \caption{}
    \label{fig:tag-sementic}
  \end{subfigure}
  \hfill 
  \begin{subfigure}{0.34\linewidth}
    \includegraphics[width=\textwidth]{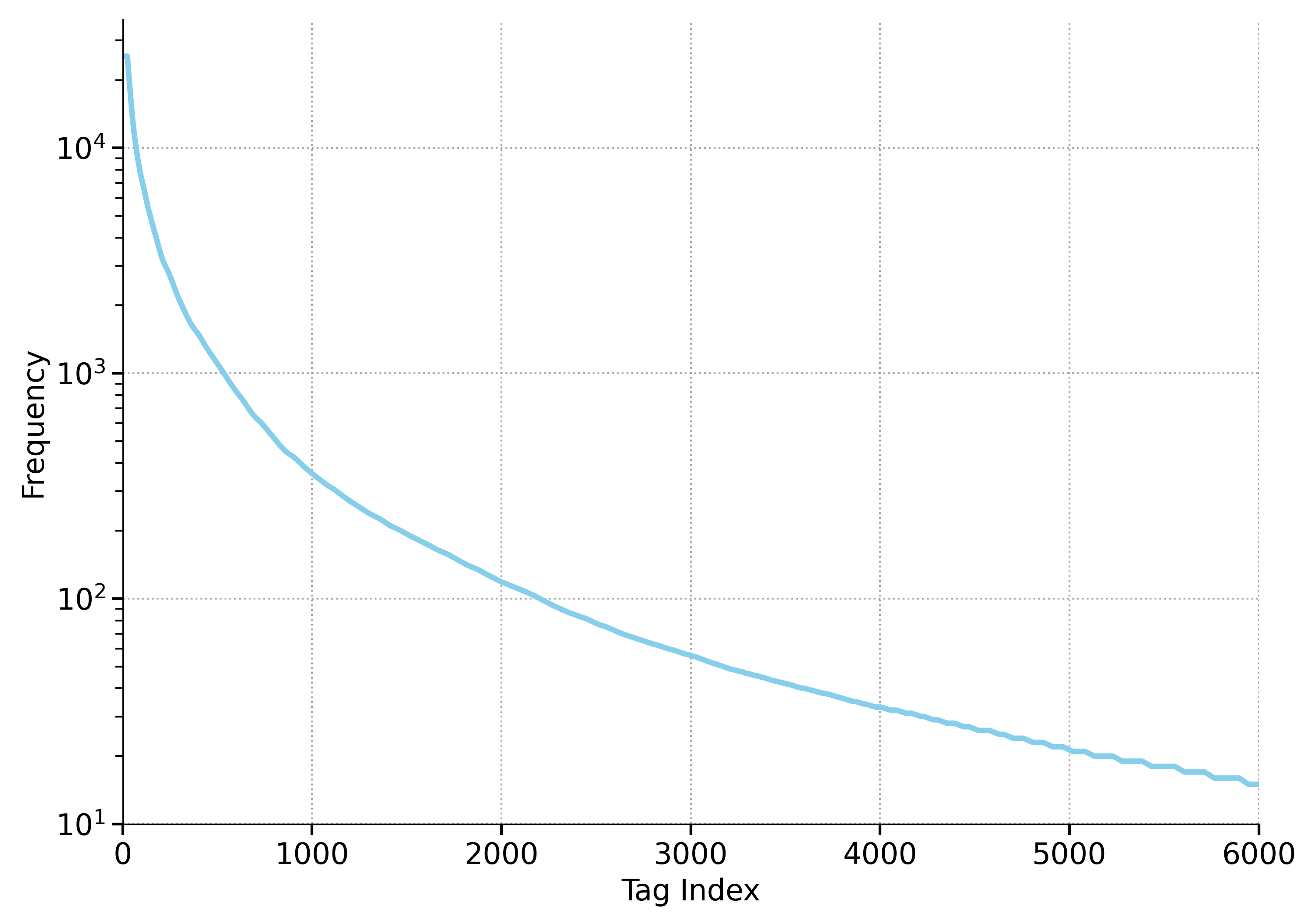}
    \caption{}
    \label{fig:frequency-plot}
  \end{subfigure}
  \caption{Analysis of our tag system. (a) Word cloud of the most frequent tags in our parsed tags, illustrating its diverse vocabulary. (b) t-SNE comparison of our UTS-1.5k (blue) and AudioSet (red), demonstrating our system's superior semantic coverage and density. (c) The characteristic long-tail tag frequency distribution of all parsed tags.}
  \label{fig:tag-system-analysis}
\end{figure*}

\subsection{Training Setup}
\paragraph{Data and Pre-processing.}
All pre-training experiments are conducted on the audio data from \textit{CaptionStew} 400K-subset. Our tag-oriented objectives (MTC, PAR) are trained on the curated tag labels, while the audio-language objectives (CLAP, Captioning) utilize the final summary paragraph of the generated captions. We process all audio signals by resampling them to 16 kHz and extracting 80-dimensional log-Mel filterbank features, computed using a 25 ms window and a 10 ms hop size. All corresponding text is tokenized using a 50k-vocabulary BPE tokenizer~\citep{lewis-etal-2020-bart}.

\paragraph{Model Architectures.}
Our audio encoder $h_{\theta}$ is the Zipformer-M architecture~\citep{yao2024zipformer}, a model we selected for its high efficiency and strong performance on long sequences. Its design features a U-Net structure with six encoder blocks, which processes features at multiple resolutions to capture both fine- and coarse-grained temporal information. Although originally developed for ASR, recent experiments~\citep{anonymous2025revisiting} have confirmed its efficacy as a highly competitive backbone for general audio classification.
The text-side architectures vary by objective. For contrastive pre-training, we employ a standard BERT-base text encoder with 12 layers and a 768-dimensional hidden state~\citep{devlin-etal-2019-bert}. For captioning objectives, we use a BART-base decoder with 6 layers and a 768-dimensional hidden state~\citep{lewis-etal-2020-bart}. We selected audio encoder with roughly twice the layer number of the text decoder to ensure comparable training throughput across objectives.

\paragraph{Configuration.}
All models were trained from scratch on 8 Tesla V100 GPUs, using a per-GPU batch size of 640 audio seconds.
We tuned several key hyperparameters for our objectives. For MTC models, we experimented with five tag system vocabularies, $K \in \{800, 1\text{k}, 1.5\text{k}, 2\text{k}, 3\text{k}\}$. For the mixed Captioning objective, we used an autoregressive-to-parallel ratio of 0.25. For the Multi-task objective, we tested $\lambda$ values of $0.2$ and $1.0$.
Training schedules varied by method. The MTC models were trained for 700k steps (approximately 12 days). All other models including Tag Parallel Decoding, CLAP, Captioning, and Multi-task were trained for 400k steps.

\subsection{Evaluation Setup}
We follow the comprehensive evaluation protocol established in~\cite{anonymous2025revisiting} to assess our pre-trained encoders. This protocol assesses models across three distinct axes, with all experiments probing representations from the encoder's final layer. To ensure a fair comparison, the audio encoder weights are kept \textbf{frozen} in all three protocols.
\begin{itemize}
    \item \textbf{Linear Probing.} To assess discriminative capabilities, we train simple linear classifiers for a diverse suite of audio tasks (event classification, sound detection, speaker, and music). We experiment with two pooling mechanisms: mean-pooling and multi-head attention pooling.
    \item \textbf{Audio-language Alignment.} We evaluate both retrieval and captioning capabilities. The retrieval setup pairs the audio encoder with a pretrained RoBERTa-base text encoder~\citep{roberta}, whereas the captioning setup utilizes a pretrained BART-base decoder~\citep{lewis-etal-2020-bart} and only finetune cross-attention layers while keeping audio encoders frozen.
    \item \textbf{Open-formed Question Answering.} We benchmark general audio understanding by connecting the frozen encoder to an LLM (\textit{Qwen2.5-7B-Instruct}~\citep{qwen2025qwen25technicalreport}) and training only a lightweight adapter on the target audio QA datasets and evaluate on corresponding track in AIR-Bench~\citep{airbench}.
\end{itemize}

\subsection{Comparative Methods}
Following the principles of broad performance benchmarking established in frameworks like SUPERB~\citep{superb} and HEAR~\citep{hear}, we first establish two primary baseline sets. To benchmark our tag-oriented pre-training methods, we compare against the standard strong supervision paradigm: our Zipformer-M architecture trained on the AudioSet~\citep{audioset}. For our audio-language pre-training methods, we compare against the \textit{Contrastive-scratch} and \textit{Captioning-scratch} trained from scratch on 400k-subset with results reported in~\cite{anonymous2025revisiting}. This is a critical comparison, as those models were trained on the exact same 400k audio subset but used original captions, providing a controlled experiment to isolate the impact of caption quality on pre-training performance.

For a broader external comparison, we also selected representative, publicly available SSL models. These baselines were chosen to span different training paradigms and specialized audio domains. For speech, we include the canonical Wav2vec 2.0~\citep{wav2vec2}, which learns by distinguishing quantized latent representations from distractors. The music domain is represented by MERT~\citep{MERT}, a model that employs masked acoustic modeling to capture structural and acoustic cues. For general-purpose audio, we use BEATs~\citep{BEATs}, a model based on an iterative masked acoustic token prediction framework. This collection of specialized and general-purpose models provides the context to assess our encoder's performance as a general-purpose audio representation.

\subsection{Dataset Analysis}
\label{sec:dataset-analysis}

\paragraph{Quantitative Comparison.} We provide a high-level comparison against prominent existing audio tagging datasets in Table.~\ref{table:tag-dataset-comparison}. We observe that most established benchmarks are either highly domain-specific (e.g., CREMA-D, MagnaTagATune) or, if general-purpose (e.g., AudioSet) are limited to a smaller and potentially biased vocabulary. In contrast, our data-driven approach yields the first large-scale, comprehensive tag system that naturally unifies all three major audio domains: speech, music, and environmental sounds. Although our data size (400k) is smaller than that of AudioSet ($\sim$2M), we hypothesize that the stronger, higher-quality supervisory signal provided by our unified tag system and caption can outweigh the larger data volume of existing benchmarks.

\paragraph{Qualitative and Statistical Analysis.} In Fig.~\ref{fig:wordcloud}, we present a word cloud of the most frequent tags via LLM parsing. This qualitatively demonstrates the vocabulary's diversity, capturing not only primary sound sources (music, male, bass, guitar) but also crucial acoustic properties (reverb, distortion, electronic) and recording environments (studio), which are often missing from human-curated taxonomies. Fig.~\ref{fig:frequency-plot} plots the tag frequency distribution, which follows the characteristic long-tail curve common to naturalistic data, validating our use of the TF-IDF algorithm to identify the most salient concepts.

\vspace{-5pt}
\paragraph{Semantic Coverage vs. AudioSet.} To quantitatively assess the semantic coverage of our UTS against the current standard, we conduct a comparative visualization in Fig.~\ref{fig:tag-sementic}. We project the semantic embeddings of our UTS (K=1.5k) and the AudioSet tag set (K=527) into a 2D space using t-SNE~\citep{tsne}. The tag embeddings are generated using the SentenceTransformer~\citep{sentencebert}. The visualization clearly shows that the semantic space occupied by AudioSet's tags (red) is fully encompassed by the much larger, denser space of our UTS (blue). This strongly suggests that our proposed tag system achieves complete semantic coverage of the primary audio benchmark, while simultaneously offering significantly greater granularity and density.

\begin{table}[t]
    \caption{Comparison of existing audio tagging datasets.}
    \vspace{-5pt}
    \label{table:tag-dataset-comparison}
    \centering
    \setlength{\tabcolsep}{5pt} 
    \renewcommand{\arraystretch}{1.0} 
    \resizebox{8.5cm}{!}{
    \small
    \begin{tabular}{l c l l} 
    \toprule
    Dataset & Audio Coverage & Tag Size & Data Size\\
    \hline
    AudioSet~\citep{audioset} & General events (incl. speech, music) & 527 &  $\sim$ 2M\\
    FSD-50k~\citep{fsd50k} &  General sound events& 200 & $\sim$ 50k\\
    VggSound~\citep{vggsound} & General sound events & 309 & $\sim$200k\\
    CREMA-D~\citep{cao2014crema} & Speech & 6 & $\sim$ 7k\\
    MagnaTagATune~\citep{MagnaTagATune} & Music & 50 & $\sim$ 24k\\
    \textbf{Ours} & Unified (Speech, Music, Env. Sounds) & 800/1k/1.5k/2k/3k & $\sim$ 400k\\
    \bottomrule
    \end{tabular}}
\end{table}

\subsection{Main Results}
We present our main results across our evaluation protocols. Our primary results for linear probing (with mean-pooling) and audio-language alignment are in Table~\ref{tab:main-table}. We provide a supplementary comparison of pooling mechanisms in Table~\ref{tab:mhap} (multi-head attention pooling), and our open-formed question answering (QA) results are in Table~\ref{tab:qa}.

\begin{table*}[t]
\centering
\renewcommand{\arraystretch}{1.1}
\caption{Main results on linear probing (with mean-pooling) and audio-language alignment. The linear probing tasks span general audio (FSD-50k~\citep{fsd50k}, VggSound~\citep{vggsound}, AudioSet-Strong~\citep{audiosetstrong}), speech (VoxCeleb2~\citep{voxceleb2}, CREMA-D~\citep{cao2014crema}), and music (MagnaTagATune~\citep{MagnaTagATune}, NSynth~\citep{NSynth}). See Supplementary Material Sec. 8 for full details of evaluation datasets and metrics.
For audio-language alignment, AC, PSC, and MC refer to the AudioCaps~\citep{audiocaps}, ParaSpeechCaps~\citep{paraspeechcaps}, and MusicCaps~\citep{MusicCaps}, respectively. We \textbf{bold} the best score among our proposed models. Baseline scores that are \underline{underlined} are surpassed by their corresponding ``Ours" counterpart. $\dagger$ denotes scores quoted from prior work with a consistent evaluation setup.}
\label{tab:main-table}
\resizebox{\textwidth}{!}{
\begin{tabular}{l l| c c c c c c c| c c c | c c c}
\hline
\multirow{3}{*}{\textbf{Methods}} & \multirow{3}{*}{\textbf{Label}} & \multicolumn{7}{c|}{\textbf{Linear Probing}} & \multicolumn{6}{c}{\textbf{Audio-language Alignment}} \\
& & & & & & & & &  \multicolumn{3}{c|}{\textit{Captioning}} & \multicolumn{3}{c}{\textit{Retrieval}} \\
 & & FSD-50k & VggSound & AS-Strong & VoxCeleb2 & CREMA-D & MTAT & NSynth & AC & PSC & MC & AC & PSC & MC \\
\hline
\textit{\textbf{SSL Models}} \\
BEATs~\citep{BEATs} & - & 0.565{$^\dagger$} & - & 0.034{$^\dagger$} & -& -& 0.400{$^\dagger$} & 75.90{$^\dagger$} & - & - & - & - & - & - \\
Wav2vec 2.0~\citep{wav2vec2} & - & 0.342{$^\dagger$} & - & - & 51.60 & 56.10 & 0.317{$^\dagger$} & 40.20{$^\dagger$} & - & - & - & - & - & - \\
MERT~\citep{MERT} & - & - & - & - & - & - & 0.402{$^\dagger$} & 72.60{$^\dagger$} & - & - & - & - & - & -  \\
\hline
\textit{\textbf{Baselines}} \\
 MTC (AudioSet) & Tag & 0.656 & 56.46 & 0.216 & \underline{18.84} & \underline{67.14} & 0.407 & 67.19 & 46.67 & \underline{45.54} & \underline{22.91} & 40.46 & \underline{49.2} & 24.6 \\
 Contrastive-scratch~\citep{anonymous2025revisiting} & Caption & 0.493 & 43.78 & \underline{0.095} & 38.63 & \underline{63.74} & \underline{0.384} & \underline{60.91} & \underline{44.50} & 45.92 & \underline{22.07} & \underline{28.73} & \underline{55.0} & \underline{19.0}   \\
 Captioning-scratch~\citep{anonymous2025revisiting} &  Caption & \underline{0.430} & \underline{39.52} & \underline{0.077} & \underline{21.95} & \underline{60.91} & 0.378 & 57.08 & \underline{43.58} & \underline{42.85} & \underline{22.62} & \underline{26.03} & \underline{49.2} & \underline{14.2} \\
 \hline
 \textit{\textbf{Our Tag-Oriented Pre-Trained Models}}\\
 MTC (Ours-UTS) & Tag* & 0.459 & 37.70 & 0.113 & 37.10 & 66.01 & 0.375 & \textbf{63.62} & 44.40 & 45.98 & 23.28 & 26.14 & 49.6 & 15.6  \\
 PAR (Ours-UTS) & Tag Sequence* & 0.433 & 39.59 & 0.121 & \textbf{38.78} & 62.47 & 0.381 & 57.91 & 44.80 & 45.66 & \textbf{23.33} & 26.76 & 49.8 & 12.2 \\
\hline
\textit{\textbf{Our Audio-Language Pre-Trained Models}}\\
Contrastive-scratch (Ours) &  Caption* & 0.445 & 40.78 & 0.105 & 33.88 & \textbf{67.29} & \textbf{0.396} & 61.40 & 44.54 & 45.73 & 22.83 & \textbf{29.66} & \textbf{55.3} & \textbf{19.8} \\
Captioning-scratch (Ours) &  Caption* & 0.439 & 39.78 & 0.087 & 29.87 & 64.74 & 0.377 & 54.25 & 45.07 & 45.81 & 22.83 & 26.24 & 50.2 & 14.4 \\
Multi-Task (Ours) &  Tag*, Caption* & \textbf{0.485} & \textbf{40.81} & \textbf{0.140} & 34.62 & 65.31 & \textbf{0.396} & 59.94 & \textbf{45.17} & \textbf{46.01} & 23.09 & 27.07 & 49.0 & 15.8 \\
\hline
\end{tabular}
}
\end{table*}

\vspace{-5pt}
\paragraph{Tag-Oriented Pre-Training vs. Baselines.}
We first analyze our tag-oriented models (MTC (Ours-UTS) and PAR(Ours-UTS)) against the primary MTC (AudioSet) baseline. On general, in-domain audio event tasks (FSD-50k, VggSound), the MTC (AudioSet) model, trained on 5x more data (2M clips), predictably sets a high bar (e.g., Table.~\ref{tab:main-table}, FSD-50k: 0.656).
However, our models demonstrate dramatic and superior out-of-domain (OOD) generalization. This is most evident in speech tasks. On VoxCeleb2 linear probing, PAR (Ours-UTS) achieves 60.97 (Table.~\ref{tab:mhap}) and MTC (Ours-UTS) achieves 37.10 (Table.~\ref{tab:main-table}), significantly outperforming the MTC (AudioSet) baseline's 58.76 and 18.84, respectively. 
We also observe strong performance from our tag-oriented models in the audio-language alignment benchmarks. On domain-specific tasks, such as music captioning on MusicCaps, our PAR (Ours-UTS) model achieves a T2A Recall@1 of 23.33, outperforming the MTC (AudioSet) baseline of 22.91.
This trend is most pronounced in complex reasoning: in Table.~\ref{tab:qa}, MTC (Ours-UTS) achieves near-perfect accuracies on several speech tasks (e.g., 92.64 on gender, 86.59 on age) and wins on Music QA (6.16), decisively crushing the MTC (AudioSet) baseline (5.61). This strongly suggests that the high quality and unified coverage of UTS provides a more potent supervisory signal than a 5x larger dataset.

\vspace{-10pt}
\paragraph{Audio-Language Pre-Training vs. Baselines.} 
We then evaluate our audio-language pre-training models, which are trained on our SOTA-generated captions (Caption*), against the~\cite{anonymous2025revisiting} baselines, which were trained on the exact same 400k audio data but used low-quality and information-sparse captions. 
Our models demonstrate a clear and consistent advantage, surpassing the ~\cite{anonymous2025revisiting} baselines on the majority of evaluation tasks, as indicated by the underlined metrics across our tables. 
This trend is especially pronounced in Table.~\ref{tab:mhap}), where Contrastive-scratch (Ours) outperforms the baseline on 4 out of 6 linear probing tasks. 
Furthermore, in the audio-language alignment benchmarks (Table.~\ref{tab:main-table}), our Captioning-scratch (Ours) model outperforms the Captioning-scratch~\citep{anonymous2025revisiting} baseline across all evaluated metrics.
This comprehensive set of results strongly indicates that the higher information quality and semantic density of our Caption* source directly translate into more robust and versatile encoders compared to models trained on original captions.

\vspace{-5pt}
\begin{table}[h]
\centering
\renewcommand{\arraystretch}{1.05}
\caption{Open-formed Question Answering (QA) Results on the AIR-Bench~\citep{airbench}. The \textit{Sound} and \textit{Music} columns report 10-point scores from open-ended questions, evaluated by GPT-4o~\citep{openai2024gpt4technicalreport}. The five scores in the \textit{Speech} column correspond to the classification accuracy on the Emotion-MELD, Emotion-IEMOCAP, Gender-MELD, Gender-common, and Age tasks, respectively.}
\vspace{-5pt}
\label{tab:qa}
\resizebox{0.49\textwidth}{!}{
\begin{tabular}{l| c c c}
\hline
\multirow{2}{*}{\textbf{Methods}}  & \multicolumn{3}{c}{\textbf{Open-formed QA}}  \\
 & Sound & Speech (emotion/gender/age) & Music \\
\hline
\textit{\textbf{Baselines}} \\
 MTC (AudioSet) & 7.01 &  47.16/\underline{25.90}/\underline{47.27}/\underline{45.06}/\underline{37.24} & \underline{5.61} \\
 Contrastive-scratch~\citep{anonymous2025revisiting} &  \underline{5.69} &  \underline{29.61}/48.79/\underline{55.25}/84.85/86.59 & \underline{5.99}  \\
 Captioning-scratch~\citep{anonymous2025revisiting} & \underline{6.25} & \underline{22.93}/37.40/70.20/78.93/40.14 & \underline{5.73}\\
 \hline
 \textit{\textbf{Our Tag-Oriented Pre-Trained Models}}\\
 MTC (Ours-UTS) & \textbf{6.68} & 39.19/\textbf{48.79}/\textbf{74.45}/\textbf{92.64}/\textbf{86.59} & \textbf{6.16}\\
 PAR (Ours-UTS) & 6.59 & 17.90/38.64/47.82/37.40/57.16 & 6.03\\
\hline
\textit{\textbf{Our Audio-Language Pre-Trained Models}}\\
Contrastive-scratch (Ours) & 5.78 & \textbf{44.08}/41.34/73.11/77.82/40.48 & 6.02 \\
Captioning-scratch (Ours) & 6.35 & 23.47/29.90/46.40/28.02/18.82 & 5.93  \\
Multi-Task (Ours) & 6.60 & 41.70/32.86/53.38/57.00/65.47 & 6.15 \\
\hline
\end{tabular}
}
\end{table}

\begin{table*}[t]
\centering
\begin{minipage}{0.62\textwidth}
    \centering
    \renewcommand{\arraystretch}{1.0}
    \resizebox{\textwidth}{!}{
    \begin{tabular}{l| c c c c c c}
    \hline
    \multirow{2}{*}{\textbf{Methods}}  & \multicolumn{6}{c}{\textbf{Linear Probing}}  \\
     & FSD-50k & VggSound & VoxCeleb2 & CREMA-D & MTAT & NSynth \\
    \hline
    \textit{\textbf{Baselines}} \\
     MTC (AudioSet) & 0.656 & 56.23 & \underline{58.76} & 72.52 & 0.405 & 74.80 \\
     Contrastive-scratch~\citep{anonymous2025revisiting} & 0.534 & 46.79 & \underline{70.18} & \underline{69.97} & \underline{0.385} & \underline{70.19}  \\
     Captioning-scratch~\citep{anonymous2025revisiting} & \underline{0.483} & 43.43 & \underline{44.51} & \underline{66.29} & \underline{0.382} & \underline{68.31}\\
     \hline
     \textit{\textbf{Our Tag-Oriented Pre-Trained Models}}\\
     MTC (Ours-UTS) & 0.478 & 40.02 & 52.52 & 70.97 & 0.376 & 69.60 \\
     PAR (Ours-UTS) & 0.489 & 43.27 & 60.97 & 69.98 & 0.388 & 68.97 \\
    \hline
     \textit{\textbf{Our Audio-Language Pre-Trained Models}}\\
    Contrastive-scratch (Ours)  & \textbf{0.514} & \textbf{45.63} & \textbf{71.63} & \textbf{72.39} & \textbf{0.401} & \textbf{71.22} \\
    Captioning-scratch (Ours) & 0.485 & 43.53 & 49.94 & 68.56 & 0.386 & 67.29   \\
    Multi-Task (Ours) & 0.503 & 43.35 & 53.40 & 70.69 & 0.399 & 68.95 \\
    \hline
    \end{tabular}
    }
    \captionof{table}{Linear probing results with multi-head attention pooling.}
    \label{tab:mhap}
\end{minipage}
\hfill 
\begin{minipage}{0.33\textwidth} 
    \centering
    \includegraphics[width=\textwidth]{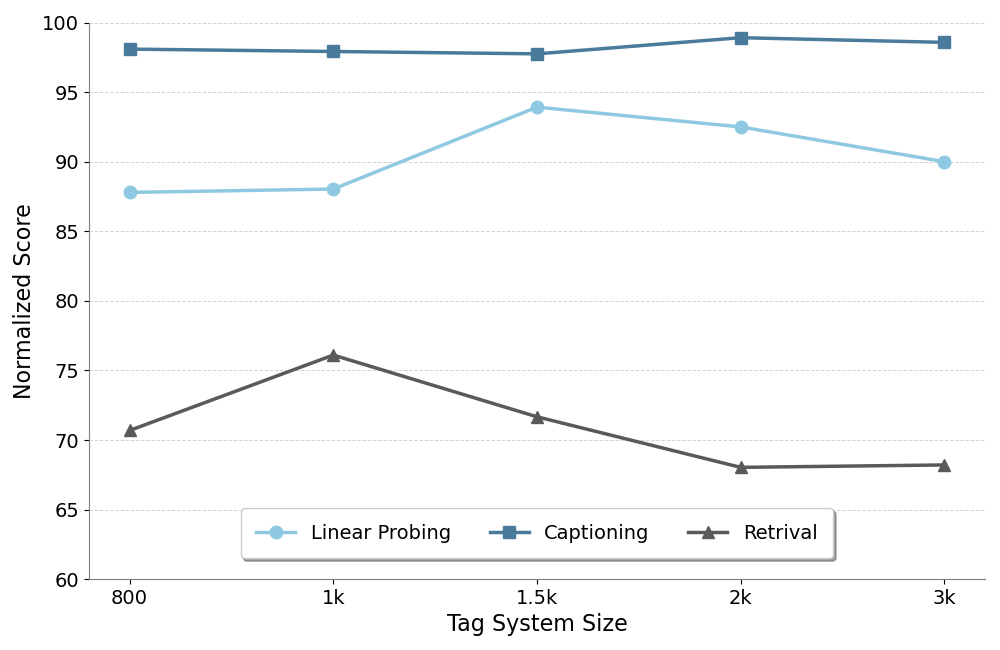} 
    \captionof{figure}{Impact of Tag System Size} 
    \label{fig:tag-num-scaling}
\end{minipage}
\end{table*}

\vspace{-8mm}
\paragraph{Analysis of Different Pre-Training Objectives.}
Our results reveal a clear specialization of objectives across the three evaluation protocols. No single method excels at all tasks; instead, the pre-training objective creates a strong inductive bias.
For Linear Probing, the choice of pooling mechanism is critical. While our Multi-Task model shows the broadest strength in mean-pooling (Table.~\ref{tab:main-table}), the Contrastive-scratch (Ours) model becomes dominant when using multi-head attention pooling (Table.~\ref{tab:mhap}), sweeping all 6 tasks. This suggests its learned features are uniquely suited to attention-based aggregation.
For Audio-Language Alignment, Contrastive-scratch (Ours) expectedly wins in retrieval. However, we observe that the Captioning-scratch (Ours) model is not the best performer on captioning tasks; this honor goes to the Multi-Task (Ours) model, suggesting that the strong MTC signal acts as a powerful regularizer for the generative objective.
Finally, for Open-formed QA, MTC (Ours-UTS) is the clear champion (Table.~\ref{tab:qa}), demonstrating that its fine-grained, strongly-supervised tag-based learning is most effective for instilling the factual, discriminative knowledge required for LLM reasoning. Conversely, the PAR model, while strong in probing, fails catastrophically in QA, highlighting its extreme task specialization.

Given these clear trade-offs, the Multi-Task (Ours) model emerges as a powerful and balanced hybrid. It achieves the best-in-class performance among all our models on general-purpose probing tasks (e.g., FSD-50k, VggSound in Table~\ref{tab:main-table}), outperforming other objectives on captioning tasks (AudioCaps, ParaSpeechCaps), and delivering strong results on Music QA (Table~\ref{tab:qa}: 6.15).
This demonstrates that combining the discrete, strong signal of our Tag* with the descriptive, rich signal of our Caption* yields a highly effective ``jack-of-all-trades" representation.

\subsection{Discussion}

\paragraph{The Impact of Tag System Size.}
We analyze the effect of our UTS vocabulary size $K$, as shown in Fig.~\ref{fig:tag-num-scaling}. We plotted the normalized average performance of our MTC models ($K$ = 800 to 3k) across three major task categories (linear probing, captioning and retrieval). 
For linear probing performance, we only use the results from mean-pooling, as its better reflects the intrinsic quality of the learned representations.
These scores are computed by first setting the MTC (AudioSet) baseline as 100\% for each individual task, and then averaging the relative percentage scores of our models within that category. 
For linear probing and retrieval, performance clearly peaks at $K$=1.5k or $K$=2k, after which it begins to plateau or decline. This suggests a trade-off: a larger $K$ provides a more granular vocabulary but also introduces data sparsity. For our 400k dataset, a vocabulary size of 1.5k-2k appears to be the ``sweet spot" that balances semantic richness and trainability. Generative tasks like Captioning appear more robust to larger $K$ values.
\paragraph{Data Quality as the Primary Driver of Performance.}
Results suggest supervision source quality is more critical to performance than pre-training objective. 
On our 400k dataset, the performance gap between our tag-oriented and audio-language models is often smaller than their collective gap against baselines. This ``data-first" principle~\citep{data-dl-era,conceptual-caption} is evident in two key comparisons: 1) our tag-oriented models achieve superior out-of-domain generalization over the MTC (AudioSet) baseline despite using a 5x less data, proving UTS coverage is more impactful than raw data volume; 2) when controlling data size, our audio-language models consistently outperform the baselines, proving label richness directly yields more robust encoders. 
This leads to our central observation: data matters most; objective choice appears to be a secondary factor, primarily leading to different task specializations. 
This provides a clear lesson for the audio pre-training community. While objective design is important, future efforts should be directed toward the supervision source itself, either by developing even larger, more comprehensive UTS or by advancing high-fidelity captioners, as rich text is itself the ultimate form of universal audio supervision. Fundamentally, this aligns with the core goal of learning: ensuring the data and its supervisory label are more closely and richly matched.

%% file: sec/conclusion.tex
\section{Limitation and Conclusion}
\vspace{-5pt}
We proposed a new data-centric paradigm for general-purpose audio pre-training, and introduced a large-scale Unified Tag System (UTS), the first of its kind to bridge speech, music, and environmental sounds. 
Our systematic study demonstrates that models trained on this ``strong source" outperform baselines on larger, non-unified datasets. This confirms that data quality is the primary driver of performance, while the choice of objective dictates downstream task specialization.
We also acknowledge limitations. Our UTS is inherently biased by its single ``teacher" captioner, and the interplay between data quality and volume at massive scales remains an open question. 
This observed task specialization also highlights a key challenge for future work: designing a single, unified objective that can excel across all downstream tasks without compromise.

%% file: sec/suppl.tex
\clearpage
\setcounter{page}{1}
\maketitlesupplementary

\section{Audio Source of Training Data}
We utilize 400k audio subset from \textit{CaptionStew}~\citep{anonymous2025revisiting}, a composite dataset aggregating 8 open-source collections to mitigate data scarcity and enhance diversity in audi pre-training.
As outlined in Table.~\ref{tab:source}, the data spans diverse acoustic domains, including environmental sounds, music, and expressive speech.

\begin{table*}[th]
\caption{Overview of the public datasets constituting CaptionStew. The table summarizes their scale, domain coverage, audio sources, and diverse captioning pipelines (from human annotation to LLM generation).}
\label{tab:source}
\resizebox{\textwidth}{!}{
\begin{tabular}{llllll}
\toprule
\textbf{Dataset}                                                      & \textbf{\#audio/\#cap}                                                                 & \textbf{Domain}                                       & \textbf{Audio source}                                                                                 & \textbf{Caption style}                                                                                                                                      & \textbf{Caption generation pipeline}                                                                                                                          \\
\midrule
\begin{tabular}[c]{@{}l@{}}AudioCaps\\ \end{tabular} & 46k/46k                                                                       & \begin{tabular}[c]{@{}l@{}}general (environmental,\\  human/animal sounds)\end{tabular} & \begin{tabular}[c]{@{}l@{}}AudioSet\\ \end{tabular}                                                                                     & \begin{tabular}[c]{@{}l@{}}Human-annotated, short description\end{tabular}                                                                                                        & crowdsourced                                                                                                                                         \\[1.0em] \midrule
\begin{tabular}[c]{@{}l@{}}Clotho \end{tabular}                                                       & 5k/25k                                                                        & environmental sounds                         & FreeSound                                                                                    & \begin{tabular}[c]{@{}l@{}}Human-annotated, short description\end{tabular}                                                                                       & crowdsourced                                                                                                                                         \\ \midrule
\begin{tabular}[c]{@{}l@{}}MusicCaps\\ \end{tabular}                                                   & 3k/3k                                                                         & music                                        & \begin{tabular}[c]{@{}l@{}}AudioSet\\ \end{tabular}                                                                         & \begin{tabular}[c]{@{}l@{}}Expert musician-written,\\multi-sentence, fine-grained description\end{tabular}                                                                                  & expert curation                                                                                                                                      \\[1.0em] \midrule
\begin{tabular}[c]{@{}l@{}}WavCaps\\ \end{tabular}                                                       & 400k/400k                                                                     & \begin{tabular}[c]{@{}l@{}}general (environmental,\\  human/animal sounds)\end{tabular} & \begin{tabular}[c]{@{}l@{}}AudioSet \\ BBC Sound Effect\\ FreeSound\\ SoundBible\end{tabular} & LLM-refined captions                                                                                 & \begin{tabular}[c]{@{}l@{}}three-stage pipeline:\\web-crawled raw descriptions\\ $\rightarrow$ ChatGPT rewrite $\rightarrow$ filtering\end{tabular} \\[1.0em] \midrule
\begin{tabular}[c]{@{}l@{}}AudioSetCaps\\ \end{tabular}                                                  & \begin{tabular}[c]{@{}l@{}}1.9M/1.9M\\ 4.0M/4.0M\\ 182k/182k\end{tabular}     & \begin{tabular}[c]{@{}l@{}}general (environmental,\\  human/animal sounds)\end{tabular} & \begin{tabular}[c]{@{}l@{}}AudioSet \\ YouTube8M\\  VggSound\\ \end{tabular}                      & \begin{tabular}[c]{@{}l@{}}LLM-generated, detailed,\\multi-sentence description \end{tabular}                                                                                                   & \begin{tabular}[c]{@{}l@{}}three-stage pipeline:\\LALM attribute extraction \\ $\rightarrow$ LLM captioning \\$\rightarrow$ CLAP-based filtering\end{tabular}   \\[1.0em] \midrule
\begin{tabular}[c]{@{}l@{}}FusionAudio\\ \end{tabular}                                    & 1.2M/1.2M                                                                     & \begin{tabular}[c]{@{}l@{}}general (environmental,\\  human/animal sounds)\end{tabular} & \begin{tabular}[c]{@{}l@{}}AudioSet\\ \end{tabular}        & \begin{tabular}[c]{@{}l@{}}LLM-augmented, multi-sentence,\\visual-enhanced description \end{tabular}                                                                                                    & \begin{tabular}[c]{@{}l@{}}multimodal context fusion \\(audio, visual, metadata)\\ + LLM captioning\end{tabular}                                       \\[1.0em] \midrule
\begin{tabular}[c]{@{}l@{}}JamendoMaxCap\\  \end{tabular}           & 360k/1.8M                                                                     & music                                        & Jamendo Platform                                                                                     & \begin{tabular}[c]{@{}l@{}}LLM-augmented, multi-sentence,\\fine-grained music description \end{tabular}                                                                                              & \begin{tabular}[c]{@{}l@{}}retrieval-based\\metadata imputation\\ + LLM captioning\end{tabular}                                                       \\[1.0em] \midrule
\begin{tabular}[c]{@{}l@{}}ParaSpeechCaps\\ \end{tabular}             & \begin{tabular}[c]{@{}l@{}}116k/116k (base)\\ 924k/924k (scaled)\end{tabular} & expressive speech                            & \begin{tabular}[c]{@{}l@{}}VoxCeleb1\\  VoxCeleb2\\  EARS\\  Expresso\\  Emilia\\ \end{tabular}     & \begin{tabular}[c]{@{}l@{}}Human-annotated/LLM-augmented,\\speaking-style description\end{tabular} & \begin{tabular}[c]{@{}l@{}}crowdsourced / \\ retrieval-based\\metadata imputation\\ + LALM captioning\end{tabular}    \\
\bottomrule
\end{tabular}
}
\end{table*}

\section{Label Construction}
As described in main text, our label construction pipeline comprises two distinct stages: audio caption generation and LLM-aided tag parsing. In this section, we provide comprehensive implementation details and hyperparameter configurations for both steps.

\subsection{Captioner Details and Examples}
We employ Qwen3-Omni-30B-A3B-Captioner~\citep{Qwen3-Omni} for audio caption generation, utilizing the vLLM engine with 8-way tensor parallelism and bfloat16 precision to ensure efficiency. The data loading pipeline is managed by Lhotse~\citep{lhotse}, which applies dynamic bucketing with a batch size of 32 to optimize GPU utilization. For generation, we use nucleus sampling with a temperature of 0.6 and top-$p$ of 0.95 to balance diversity and coherence. We provide samples of the generated captions and corresponding audio files in the supplementary material.

\subsection{LLM Parser}
We employ Qwen2.5-7B-Instruct~\citep{qwen2025qwen25technicalreport} as a semantic parser to extract structured tags from unstructured audio captions. 
Utilizing vLLM with 4-way tensor parallelism, we prompt the model to generate 5–10 concise, one-word labels per sample in JSON format. The instruction ensures coverage of key acoustic dimensions—including scene, sound source, and event type—using sampling parameters of $T=0.3$ and $p=0.9$. We provided the full text of the instruction prompt.

\begin{figure*}[h]
\centering
\label{prompt}
\begin{tcolorbox}[colback=gray!10, colframe=black!50, arc=3mm, boxrule=0.5pt]
\fontsize{7.5pt}{8.5pt}\selectfont
\ttfamily 

\textbf{Prompt:}
\medskip 

You are a labeling assistant for audio descriptions.
Extract essential, compact English labels for the audio caption.\\

Output rules:\\
- Output JSON only with a single key "labels": a list of 10 labels.\\
- Each label must be ONE WORD.\\
- Return ONLY a raw JSON object with key "labels". Do not use code fences or any extra text.\\
- Use lowercase.\\
- No full sentences, no punctuation beyond hyphens.\\
- Prefer canonical, domain-relevant terms with high reusability across captions.\\
- Avoid synonyms: collapse to a single consistent term set (e.g., always use "car" not "automobile").\\
- Avoid rare or obscure words; choose common, widely recognized terms.\\
- Cover five aspects when possible: scene/environment, sound sources, sound types/events, audio/production traits, semantic function/intent.\\
- Avoid duplicates.\\

Expected JSON:\\
{{"labels": ["xxx", "xxx", "xxx", "xxx", "xxx"]}}

Now process the following caption and return JSON only:\\
CAPTION:\\
\{caption\}

\end{tcolorbox}
\end{figure*}

\section{Evaluation Details}

Table.~\ref{tab:dataset_full} provides a comprehensive breakdown of the datasets and metrics used for evaluate audio representation quality. The assessment covers three distinct protocols: linear probing, audio-language alignment and open-form question answering.

\begin{table*}[!h]
    \centering
    \caption{Overview of the evaluation datasets used for assessing audio representation quality. $^\dagger$Evaluated by GPT-4o following the AIR-Bench protocol. $^\ddagger$Synthesized from publicly available speech datasets~\citep{ardila2019common, busso2008iemocap, cao2014crema, ravdess, poria2018meld} using fixed question templates.}
    \label{tab:dataset_full}
    \resizebox{0.9\textwidth}{!}{
\begin{tabular}{llccccl}
\toprule  
Evaluation Dataset & Task   & \#samples & \#class & train  & eval                  & Metrics \\
\midrule
FSD-50k            & Multi-label audio event classification & 37,168 / 10,231 & 200 & $\checkmark$ & $\checkmark$ & mAP     \\
VggSound           & Single-label audio event classification & 183,730 / 15,446 & 309 & $\checkmark$ & $\checkmark$ & accuracy    \\
VoxCeleb2          & Speaker identification        & 1,092,009 / 36,693 & 5,994 & $\checkmark$ & $\checkmark$           & accuracy    \\
CREMA-D            & Speech emotion recognition    & 6,030 / 706 & 6 & $\checkmark$ & $\checkmark$                   & accuracy   \\
MagnaTagATune      & Music tagging           & 19,425 / 4,856 & 50 & $\checkmark$ & $\checkmark$                            & mAP     \\
NSynth             & Musical instrument classification   & 289,205 / 4,096 & 11 & $\checkmark$ & $\checkmark$                       & accuracy  \\
AudioSet-strong    & Sound event detection & 103,463 / 16,996 & 456 & $\checkmark$ & $\checkmark$     & PSDS1 \\
\midrule
AudioCaps             & \multirow{3}{*}{\begin{tabular}[c]{@{}l@{}}Text-to-audio retrieval\\ Audio captioning\end{tabular}}  & 49,838 / 975 & -- & $\checkmark$ & $\checkmark$     & \multirow{3}{*}{\begin{tabular}[c]{@{}l@{}}Recall@1\\ RougeL\end{tabular}}   \\
ParaSpeechCaps          &   &    116,516 / 500 & -- & $\checkmark$ & $\checkmark$          &    \\
MusicCaps             &    &    2,663 / 500 & -- & $\checkmark$ & $\checkmark$            &    \\
\midrule
ClothoAQA & \multirow{8}{*}{Open-formed question answering} &    7,044 & -- & $\checkmark$ & $\times$          & \multirow{8}{*}{Score$^\dagger$}\\
In-house SpeechQA$^\ddagger$ & &  160,000 & -- & $\checkmark$ & $\times$       & \\
MusicQA & & 70,011 & -- & $\checkmark$ & $\times$          & \\
AIRBench-chat-sound & &  400 & -- & $\times$ & $\checkmark$       & \\
AIRBench-foundation-emotion & &  1,000 & -- & $\times$ & $\checkmark$       & \\
AIRBench-foundation-gender & &  1,000 & -- & $\times$ & $\checkmark$       & \\
AIRBench-foundation-age & &  1,000 & -- & $\times$ & $\checkmark$       & \\
\bottomrule
\end{tabular}
}
    
    \label{tab:placeholder}
\end{table*}

\section{Complete Results}
We have the complete results on three evaluation axes in Table.~\ref{tab:main-table-com} and Table.~\ref{tab:mhap-com}.
The main additions are the results for five values of $K$ ranging from 800 to 3k, as well as the results for the multi-task setting where $\lambda$ is set to 0.25 and 1, respectively.

\begin{table*}[t]
\centering
\renewcommand{\arraystretch}{1.1}
\caption{\textbf{Complete} results on linear probing (with mean-pooling) and audio-language alignment. The linear probing tasks span general audio (FSD-50k~\citep{fsd50k}, VggSound~\citep{vggsound}, AudioSet-Strong~\citep{audiosetstrong}), speech (VoxCeleb2~\citep{voxceleb2}, CREMA-D~\citep{cao2014crema}), and music (MagnaTagATune~\citep{MagnaTagATune}, NSynth~\citep{NSynth}). For audio-language alignment, AC, PSC, and MC refer to the AudioCaps~\citep{audiocaps}, ParaSpeechCaps~\citep{paraspeechcaps}, and MusicCaps~\citep{MusicCaps}, respectively. We \textbf{bold} the best score in our tag-oriented models and audio-language models respectively. Baseline scores that are \underline{underlined} are surpassed by their corresponding ``Ours" counterpart. $\dagger$ denotes scores quoted from prior work with a consistent evaluation setup.}
\label{tab:main-table-com}
\resizebox{\textwidth}{!}{
\begin{tabular}{l l| c c c c c c c| c c c | c c c}
\hline
\multirow{3}{*}{\textbf{Methods}} & \multirow{3}{*}{\textbf{Label}} & \multicolumn{7}{c|}{\textbf{Linear Probing}} & \multicolumn{6}{c}{\textbf{Audio-language Alignment}} \\
& & & & & & & & &  \multicolumn{3}{c|}{\textit{Captioning}} & \multicolumn{3}{c}{\textit{Retrieval}} \\
 & & FSD-50k & VggSound & AS-Strong & VoxCeleb2 & CREMA-D & MTAT & NSynth & AC & PSC & MC & AC & PSC & MC \\
\hline
\textit{\textbf{SSL Models}} \\
BEATs~\citep{BEATs} & - & 0.565{$^\dagger$} & - & 0.034{$^\dagger$} & -& -& 0.400{$^\dagger$} & 75.90{$^\dagger$} & - & - & - & - & - & - \\
Wav2vec 2.0~\citep{wav2vec2} & - & 0.342{$^\dagger$} & - & - & 51.60 & 56.10 & 0.317{$^\dagger$} & 40.20{$^\dagger$} & - & - & - & - & - & - \\
MERT~\citep{MERT} & - & - & - & - & - & - & 0.402{$^\dagger$} & 72.60{$^\dagger$} & - & - & - & - & - & -  \\
\hline
\textit{\textbf{Baselines}} \\
 MTC (AudioSet) & Tag & 0.656 & 56.46 & 0.216 & \underline{18.84} & \underline{67.14} & 0.407 & 67.19 & 46.67 & \underline{45.54} & \underline{22.91} & 40.46 & \underline{49.2} & 24.6 \\
 Contrastive-scratch~\citep{anonymous2025revisiting} & Caption & 0.493 & 43.78 & \underline{0.095} & 38.63 & \underline{63.74} & \underline{0.384} & \underline{60.91} & \underline{44.50} & 45.92 & \underline{22.07} & \underline{28.73} & \underline{55.0} & \underline{19.0}   \\
 Captioning-scratch~\citep{anonymous2025revisiting} &  Caption & \underline{0.430} & \underline{39.52} & \underline{0.077} & \underline{21.95} & \underline{60.91} & 0.378 & 57.08 & \underline{43.58} & \underline{42.85} & \underline{22.62} & \underline{26.03} & \underline{49.2} & \underline{14.2} \\
 \hline
 \textit{\textbf{Our Tag-Oriented Pre-Trained Models}}\\
 MTC (Ours-UTS, $K$=800) & Tag* & 0.448 & 37.01 & 0.095 & 30.04 & 61.34 & 0.375 & 62.84 & 44.40 & 45.72 & 22.62 & 23.96 & 48.4 & 13.4   \\
 MTC (Ours-UTS, $K$=1k) & Tag* & 0.455 & 36.79 & 0.095 & 30.26 & 62.90 & 0.373 & 61.50 & 44.20 & \textbf{45.98} & 22.47 & 25.93 & 49.6 & \textbf{15.6}  \\
 MTC (Ours-UTS, $K$=1.5k) & Tag*  & \textbf{0.459} & 37.70 & 0.104 & 37.10 & 64.31 & 0.368 & 60.01 & 43.66 & 45.50 & 22.86 & 26.41 & 48.4 & 12.8\\
 MTC (Ours-UTS, $K$=2k) & Tag* & 0.450 & 37.48 & 0.113 & 33.22 & \textbf{66.01} & 0.370 & \textbf{63.62} & 44.06 & 45.86 & 23.28 & 24.69 & 46.8 & 11.8 \\
 MTC (Ours-UTS, $K$=3k) & Tag* & 0.449 & 37.68 & 0.113 & 30.63 & 65.02 & 0.373 & 61.57 & 44.09 & 45.87 & 23.03 & 24.90 & 46.0 & 12.2 \\
 PAR (Ours-UTS) & Tag Sequence* & 0.433 & \textbf{39.59} & \textbf{0.121} & \textbf{38.78} & 62.47 & \textbf{0.381} & 57.91 & \textbf{44.80} & 45.66 & \textbf{23.33} & \textbf{26.76} & \textbf{49.8} & 12.2 \\
\hline
\textit{\textbf{Our Audio-Language Pre-Trained Models}}\\
Contrastive-scratch (Ours) &  Caption* & 0.445 & 40.78 & 0.105 & 33.88 & \textbf{67.29} & \textbf{0.396} & \textbf{61.40} & 44.54 & 45.73 & 22.83 & \textbf{29.66} & \textbf{55.3} & \textbf{19.8} \\
Captioning-scratch (Ours) &  Caption* & 0.439 & 39.78 & 0.087 & 29.87 & 64.74 & 0.377 & 54.25 & 45.07 & 45.81 & 22.83 & 26.24 & 50.2 & 14.4 \\
Multi-Task (Ours, $\lambda=0.2$) &  Tag*, Caption* & 0.468 & 39.47 & 0.130 & \textbf{34.62} & 62.61 & 0.386 & 57.71 & 44.88 & \textbf{46.01} & \textbf{23.20} & 25.21 & 49.0 & 13.5  \\
Multi-Task (Ours, $\lambda=1$) &  Tag*, Caption* &  \textbf{0.485} & \textbf{40.81} & \textbf{0.140} & 33.12 & 65.31 & \textbf{0.396} & 59.94 & \textbf{45.17} & 45.55 & 23.09 & 27.07 & 47.0 & 15.8\\
\hline
\end{tabular}
}
\end{table*}

\begin{table*}[t]
\centering
\captionof{table}{\textbf{Complete} results on linear probing (with multi-head attention pooling) and open-formed QA. The \textit{Sound} and \textit{Music} columns report 10-point scores on open-ended questions from AIR-Bench, evaluated by GPT-4o~\citep{openai2024gpt4technicalreport}. The five scores in the \textit{Speech} column correspond to the classification accuracy on the Emotion-MELD, Emotion-IEMOCAP, Gender-MELD, Gender-common, and Age tasks, respectively.}
\renewcommand{\arraystretch}{1.1}
    \resizebox{\textwidth}{!}{
    \begin{tabular}{l| c c c c c c|c c c}
    \hline
    \multirow{2}{*}{\textbf{Methods}}  & \multicolumn{6}{c|}{\textbf{Linear Probing}}  & \multicolumn{3}{c}{\textbf{Open-formed QA}} \\
     & FSD-50k & VggSound & VoxCeleb2 & CREMA-D & MTAT & NSynth & Sound & Speech (emotion/gender/age) & Music \\
    \hline
    \textit{\textbf{Baselines}} \\
     MTC (AudioSet) & 0.656 & 56.23 & \underline{58.76} & 72.52 & 0.405 & 74.80 & 7.01 & 47.16/\underline{25.90}/\underline{47.27}/\underline{45.06}/\underline{37.24} & \underline{5.61}\\
     Contrastive-scratch~\citep{anonymous2025revisiting} & 0.534 & 46.79 & \underline{70.18} & \underline{69.97} & \underline{0.385} & \underline{70.19} & \underline{5.69} &  \underline{29.61}/48.79/\underline{55.25}/84.85/86.59 & \underline{5.99} \\
     Captioning-scratch~\citep{anonymous2025revisiting} & \underline{0.483} & 43.43 & \underline{44.51} & \underline{66.29} & \underline{0.382} & \underline{68.31} & \underline{6.25} & \underline{22.93}/37.40/70.20/78.93/40.14 & \underline{5.73} \\
     \hline
     \textit{\textbf{Our Tag-Oriented Pre-Trained Models}}\\
     MTC (Ours-UTS, $K$=800) & 0.469 & 39.56 & 45.98 & 67.57 & 0.376 & 68.97 & 6.54 & 29.61/\textbf{48.79}/55.25/84.85/\textbf{86.59} & 5.99\\
     MTC (Ours-UTS, $K$=1k) & 0.472 & 39.84 & 46.02 & 68.71 & 0.376 & 69.56  & 6.46 & 18.45/30.10/48.91/56.35/58.66 & 5.77 \\
     MTC (Ours-UTS, $K$=1.5k) & 0.477 & 39.94 & 52.52 & 69.41 & 0.371 & \textbf{69.60} & 6.48 & \textbf{39.19}/32.10/62.66/82.06/56.26 & 5.80 \\
     MTC (Ours-UTS, $K$=2k) & 0.478 & 39.78 & 48.13 & \textbf{70.97} & 0.371 & 68.92  & 6.47 & 26.20/32.60/54.15/73.49/40.84 & \textbf{6.16}\\
     MTC (Ours-UTS, $K$=3k) & 0.471 & 40.02 & 45.11 & 69.29 & 0.371 & 69.41 & \textbf{6.68} & 22.38/29.70/\textbf{74.45}/\textbf{92.64}/63.06 & 5.95\\
     PAR (Ours-UTS) & \textbf{0.489} & \textbf{43.27} & \textbf{60.97} & 69.98 & \textbf{0.388} & 68.97 & 6.59 & 17.90/38.64/47.82/37.40/57.16 & 6.03 \\
    \hline
     \textit{\textbf{Our Audio-Language Pre-Trained Models}}\\
    Contrastive-scratch (Ours)  & \textbf{0.514} & \textbf{45.63} & \textbf{71.63} & \textbf{72.39} & \textbf{0.401} & \textbf{71.22} & 5.78 & \textbf{44.08}/\textbf{41.34}/\textbf{73.11}/\textbf{77.82}/40.48 & 6.02 \\
    Captioning-scratch (Ours) & 0.485 & 43.53 & 49.94 & 68.56 & 0.386 & 67.29 & 6.35 & 23.47/29.90/46.40/28.02/18.82 & 5.93  \\
    Multi-Task (Ours, $\lambda=0.25$) & 0.490 & 41.99 & 50.80 & 69.70 & 0.399 & 67.33 & \textbf{6.60} & 41.70/26.90/57.10/53.02/45.15 & 6.09 \\
    Multi-Task (Ours, $\lambda=1$) & 0.503 &  43.35 & 53.4 & 70.69 & 0.393 & 68.95 & 6.38 & 24.62/32.86/53.38/57.00/\textbf{65.47} & \textbf{6.15} \\
    \hline
    \end{tabular}
    }
    \label{tab:mhap-com}
\end{table*}

\input{sec/samples}

%% file: sec/samples.tex
\begin{figure*}[t] 
    \centering
    \label{caption-example}
    \fbox{
        \begin{minipage}{0.9\textwidth}
            
            \textbf{Audio id: Y7IOszZm4n\_I}\\

            The audio begins in a quiet, indoor environment, marked by a persistent low-frequency hum and a faint high-frequency hiss, likely produced by an appliance or HVAC system. The atmosphere is calm and still, with no background speech or music. A soft, low-pitched rustling sound, characteristic of a dog's movement through dry leaves or grass, is heard near the recording device, followed by a subtle, low-frequency thump suggesting the dog's body settling or shifting weight. Shortly after, a faint, high-pitched squeak—possibly from a toy or collar—adds to the sense of a dog's presence. A sharp, brief rustling indicates the dog's claws or paws moving across a rough surface.\\
            Suddenly, a male voice, calm and authoritative, commands, “Come.” His speech is clear, mid-to-low in pitch, and delivered with a neutral General American accent, indicating a confident and focused tone. Immediately following the command, a sharp, percussive tongue click is produced, serving as a non-verbal cue to reinforce the verbal command. The dog responds with a distinct, high-pitched bark, signaling acknowledgment and readiness. The dog’s movement continues with a pronounced rustling sound as it shifts or stands up, accompanied by the jingle of metal tags on its collar.\\
            The male then issues the command, “Drop,” in a firm and slightly more urgent tone, indicating an expectation of a specific action. The dog’s response is immediate and enthusiastic: a rapid burst of barks and excited yips, accompanied by continued rustling and the jingling of tags as it moves energetically, possibly jumping or turning. The male repeats the command, “Drop,” with increased urgency, reflecting his expectation for the dog to comply. The dog’s barking escalates in intensity and pitch, now sounding frantic and excited, while the rustling and jingling persist. The clip ends abruptly, with all sounds ceasing at once, leaving no fade-out or lingering noise.\\
            Throughout the recording, the audio quality is moderate to low, with clear speech but noticeable hiss and hum, and the environment appears to be a small to medium-sized, enclosed indoor space with hard surfaces. The dog’s vocalizations and movements are prominent and close to the microphone, indicating the device is positioned near both the dog and the handler. The interaction is structured and purposeful, with the handler issuing commands and the dog responding with excitement and obedience. The context suggests a focused training or play session between a handler and a small or medium-sized, energetic dog, typical of North American pet ownership and training practices.\\
            In summary, this audio clip documents a succinct, indoor training exchange between a man and his dog. The man gives the commands “Come” and “Drop,” reinforced by a tongue click, and the dog responds with barks, yips, rustling, and tag jingling. The setting is a quiet, enclosed space with a steady background hum and hiss, and the interaction exemplifies a structured, purposeful moment of communication and obedience training typical of North American pet culture.
            \\
            \\
            \textbf{Tags}: [``indoors", ``appliance", ``rustling", ``dog", ``squeak", ``command", ``click", ``bark", ``training", ``hiss"]

        \end{minipage}
    }
\end{figure*}

\begin{figure*}[t] 
    \centering
    \fbox{
        \begin{minipage}{0.9\textwidth}
            
            \textbf{Audio id: MG0ThC4TSoQ\_94\_10}\\

            The audio clip begins with a burst of energetic, high-fidelity electronic music characteristic of late 1980s to early 1990s Chicago House and Acid House. The soundscape is dominated by a punchy four-on-the-floor kick drum, sharp synthetic claps or snares, and a crisp hi-hat pattern, all tightly sequenced to drive the rhythm. A deep, resonant synth bassline—likely from a Roland TB-303 or similar—delivers the iconic acid squelch, while a bright, syncopated arpeggiated synthesizer line adds melodic complexity. A recurring, high-pitched, and slightly distorted vocal sample of a female voice saying ``for you" is chopped and looped, functioning as both a rhythmic and melodic hook. The stereo image is expansive, with drums and bass centered, synth layers panned for width, and vocal samples and percussion spread across the field. The overall production is clean, loud, and compressed, with no ambient noise or room tone, suggesting a studio or digital origin.\\
            At the five-second mark, the intensity peaks as the bass and synth layers momentarily drop out, leaving the drums and a sharp snare-like clap isolated. This creates a brief, suspenseful moment of anticipation before the full instrumental arrangement resumes at full volume. The acid bassline and arpeggiated synth return, and a new vocal sample enters: a processed male voice, heavily treated with reverb and echo, delivers the phrase ``if it's good to you." The vocal is delivered in a rhythmic, spoken-word style, with the word \"good\" notably emphasized and repeated, creating a hypnotic effect. The phrase is clipped to fit the musical rhythm and is repeated for emphasis, with no additional lyrics or dialogue.\\
            As the track reaches its conclusion, all elements—drums, bass, synths, and vocals—abruptly cut off, leaving no fade-out or lingering sound. The sudden silence suggests the clip is an excerpt from a longer, continuous musical work.\\
            In summary, this audio is a meticulously produced, high-energy excerpt from a Chicago House/Acid House track, distinguished by its classic drum machine patterns, acid bassline, arpeggiated synth hooks, and iconic vocal samples. The production is studio-clean, with deliberate stereo placement and effects, and the clip serves as a vivid snapshot of early rave culture, evoking nostalgia for the golden age of electronic dance music.
            \\
            \\
            \textbf{Tags}: [``music", ``electronic", ``house", ``acid", ``drum", ``bass", ``synth", ``vocal", ``production", ``nostalgia"]

        \end{minipage}
    }
\end{figure*}

\begin{figure*}[t] 
    \centering
    \fbox{
        \begin{minipage}{0.9\textwidth}
            
            \textbf{Audio id: YITcnYeETGUE}\\

            The audio clip begins with a faint rustle, likely from the speaker adjusting a microphone or clothing, immediately followed by a soft inhalation, indicating the speaker is preparing to address the audience. The recording is marked by a persistent, low-level hiss from the microphone’s noise floor, and a subtle, low-frequency hum typical of a small, untreated indoor space. A single male speaker, with a clear, slightly nasal Italian voice and a regional accent from Northern or Central Italy, speaks in a calm, measured, and friendly manner. He uses a conversational tone, employing filler words and informal constructions, which suggests he is addressing a general audience in a relaxed, personal style.\\
            The speaker’s speech starts abruptly, mid-sentence: “…praticamente dicevo un treno Shinkansen l’ottavo giorno, andate verso Hiroshima. Appena arrivati a Hiroshima, e visitate il parco della pace, vicino al quale potete prendere un hotel…” (“…practically, I was saying, a Shinkansen train on the eighth day, go towards Hiroshima. As soon as you arrive in Hiroshima, and visit the Peace Park, near which you can take a hotel…”). The language is Italian with English loanwords (“Shinkansen”), reflecting modern Italian usage and a travel-oriented context. The speaker references the eighth day of a trip, the Shinkansen high-speed train, the city of Hiroshima, and the Peace Park, indicating he is providing travel advice or narrating a personal itinerary. His tone remains neutral and friendly, with no emotional emphasis or change in delivery. The recording ends abruptly, cutting off the final word “hotel” and leaving the speech unfinished, which suggests an accidental or intentional truncation.\\
            Throughout the clip, the audio quality is moderate, with clear speech but some muffling due to the acoustics of the small room. The frequency response is limited, lacking deep bass and crisp treble, and the signal is free from clipping or distortion. The speaker’s voice is close-mic’d, and there are no other voices, background noises, or music. The content and style indicate that the speaker is likely a young-to-middle-aged Italian male sharing travel advice or narrating a personal journey, aimed at an audience interested in Japanese travel, with a focus on Hiroshima and its Peace Park. The use of Italian with English loanwords and the informal, conversational approach suggest a personal or semi-professional travel vlog or podcast.\\
            In summary, the audio captures a brief, unfinished segment of a single male Italian speaker providing travel advice, specifically mentioning the eighth day of a trip to Hiroshima via Shinkansen, the visit to the Peace Park, and nearby hotel options. The recording is clear and conversational, set in a small indoor space with mild background hiss and hum, and ends abruptly mid-word, reflecting a personal travel narrative intended for a general audience.
            \\
            \\
            \textbf{Tags}: [``speech", ``adjustment", ``inhale", ``hiss", ``hum", ``italian", ``calm", ``conversational", ``travel", ``abrupt"]

        \end{minipage}
    }
\end{figure*}

\begin{figure*}[t] 
    \centering
    \fbox{
        \begin{minipage}{0.9\textwidth}
            
            \textbf{Audio id: 963530\_0}\\

            The audio clip is a 30-second, low-fidelity, stereo field recording featuring a solo acoustic guitar performance. The guitar is captured in a very intimate, close-mic’d manner, revealing not only the instrument’s full-bodied sound but also every nuance of the performer’s technique: finger slides, string squeaks, and the tactile sounds of the performer adjusting their hand and body.\\
            The performance is structured in three clear, non-repetitive sections. It begins with a slow, bluesy riff in a minor key, played in a fingerstyle manner. The second section introduces a more rhythmic, percussive strumming pattern, while the third section features a melodic, arpeggiated passage. Throughout, the playing is expressive and nuanced, with deliberate dynamics and a sense of improvisation. The musical style, including the choice of chords, fingerstyle technique, and melodic phrasing, is strongly rooted in the American folk-blues tradition, reminiscent of early 20th-century rural or country blues, but presented with modern recording clarity.\\
            The recording environment is outdoors, as evidenced by a constant, low-frequency wind rumble and occasional rustling of clothing, but there are no other environmental or human sounds. The absence of any reverb, echo, or ambient noise points to an open, rural or semi-rural location, likely away from urban or natural sound sources.\\
            The technical quality of the recording is compromised by a persistent hiss and wind rumble, and the guitar’s resonance is slightly muffled, indicating the use of a consumer-grade recording device. The stereo field is narrow, with the guitar and ambient sounds centered, and the overall sound lacks the brightness and detail of a professional studio recording.\\
            The clip ends abruptly, with a sharp click indicating the recording was manually stopped. Immediately after, a short, low-frequency electronic tone is heard, possibly a notification or system sound from the recording device, marking the end of the session.\\
            In summary, the audio is a raw, unedited field recording of a solo acoustic guitar performance in a wind-swept outdoor setting, capturing the essence of American folk-blues tradition in a modern, intimate, and slightly lo-fi context. The performer’s technical skill and expressive intent are clear, but the recording’s limitations and environmental noise contribute to a sense of authenticity and immediacy, evoking a solitary, reflective musical moment in a rural landscape.
            \\
            \\
            \textbf{Tags}: [``outdoor", ``guitar", ``blues", ``intimate", ``low\_fidelity", ``field\_recording", ``folk", ``acoustic", ``expressive", ``natural\_noise"]

        \end{minipage}
    }
\end{figure*}

\begin{figure*}[t] 
    \centering
    \fbox{
        \begin{minipage}{0.9\textwidth}
            
            \textbf{Audio id: 71xd4wFkozw\_62\_10}\\

            The audio clip opens with a male commentator, speaking in a clear, urgent, and slightly raised North American accent, his voice reverberating through a large, echoic space. He poses a rhetorical question: “Why would he let his mouth get into the point of swelling up that much?” His delivery is fast-paced and marked by rising intonation, reflecting tension and disbelief at the apparent risk involved in a physical altercation. The speech is accompanied by a persistent, low-level hiss, typical of analog tape recordings from the late 20th century, and is further colored by the room’s acoustics, which impart a hollow echo to the commentator’s voice.\\
            As the commentator speaks, the soundscape intensifies with a series of sharp, percussive impacts—dry, hollow thuds that suggest blows landing on flesh, particularly on the head or face, and are interlaced with guttural grunts and strained vocalizations. These non-verbal sounds, including a forceful grunt around the midpoint, indicate the physical exertion and pain of combatants, likely engaged in a martial arts or boxing match. The impacts and vocalizations reverberate through the same spacious, hard-walled venue, with the microphone capturing both direct and ambient sound.\\
            A female voice soon enters, her tone urgent and high-pitched, calling out “Yeah!” and “Come on!” in quick succession. Her shouts, delivered with the same reverberant qualities as the commentator’s, serve as a rallying cry or encouragement, possibly directed at the fighters or the audience. Her brief interjections are marked by clarity and emotional intensity, cutting through the ongoing commotion.\\
            The recording ends abruptly with the female’s last exclamation, leaving the fight unresolved and the soundscape abruptly silenced, a sign of analog tape’s limitations.\\
            In summary, this analog recording captures a moment of intense physical competition within a large, reverberant venue, likely a gymnasium or arena, during a martial arts or boxing match. The male commentator’s urgent rhetorical question and the female’s supportive shouts frame the ongoing struggle, which is punctuated by realistic impact sounds and strained vocalizations. The technical imperfections of the recording—hiss, echo, and abrupt ending—reinforce its authenticity and place it firmly within the context of late 20th-century North American sporting events.
            \\
            \\
            \textbf{Tags}: [``commentator", ``male", ``urgent", ``hiss", ``impact", ``grunts", ``female", ``encouragement", ``venue", ``boxing"]

        \end{minipage}
    }
\end{figure*}

%% file: main.bbl
\begin{thebibliography}{98}
\providecommand{\natexlab}[1]{#1}
\providecommand{\url}[1]{\texttt{#1}}
\expandafter\ifx\csname urlstyle\endcsname\relax
  \providecommand{\doi}[1]{doi: #1}\else
  \providecommand{\doi}{doi: \begingroup \urlstyle{rm}\Url}\fi

\bibitem[Agostinelli et~al.(2023)Agostinelli, Denk, Borsos, Engel, Verzetti, Caillon, Huang, Jansen, Roberts, Tagliasacchi, et~al.]{MusicCaps}
Andrea Agostinelli, Timo~I Denk, Zal{\'a}n Borsos, Jesse Engel, Mauro Verzetti, Antoine Caillon, Qingqing Huang, Aren Jansen, Adam Roberts, Marco Tagliasacchi, et~al.
\newblock Musiclm: Generating music from text.
\newblock \emph{arXiv preprint arXiv:2301.11325}, 2023.

\bibitem[Ardila et~al.(2019)Ardila, Branson, Davis, Henretty, Kohler, Meyer, Morais, Saunders, Tyers, and Weber]{ardila2019common}
Rosana Ardila, Megan Branson, Kelly Davis, Michael Henretty, Michael Kohler, Josh Meyer, Reuben Morais, Lindsay Saunders, Francis~M Tyers, and Gregor Weber.
\newblock Common voice: A massively-multilingual speech corpus.
\newblock \emph{arXiv preprint arXiv:1912.06670}, 2019.

\bibitem[Baevski et~al.(2020)Baevski, Zhou, Mohamed, and Auli]{wav2vec2}
Alexei Baevski, Yuhao Zhou, Abdelrahman Mohamed, and Michael Auli.
\newblock wav2vec 2.0: A framework for self-supervised learning of speech representations.
\newblock \emph{Advances in neural information processing systems}, 33:\penalty0 12449--12460, 2020.

\bibitem[Baevski et~al.(2022)Baevski, Hsu, Xu, Babu, Gu, and Auli]{data2vec}
Alexei Baevski, Wei-Ning Hsu, Qiantong Xu, Arun Babu, Jiatao Gu, and Michael Auli.
\newblock Data2vec: A general framework for self-supervised learning in speech, vision and language.
\newblock In \emph{International conference on machine learning}, pages 1298--1312. PMLR, 2022.

\bibitem[Bai et~al.(2025)Bai, Liu, Wang, Shi, Wang, Plumbley, Gan, and Chen]{audiosetcaps}
Jisheng Bai, Haohe Liu, Mou Wang, Dongyuan Shi, Wenwu Wang, Mark~D Plumbley, Woon-Seng Gan, and Jianfeng Chen.
\newblock Audiosetcaps: An enriched audio-caption dataset using automated generation pipeline with large audio and language models.
\newblock \emph{IEEE Transactions on Audio, Speech and Language Processing}, 2025.

\bibitem[Bird and Loper(2004)]{bird-loper-2004-nltk}
Steven Bird and Edward Loper.
\newblock {NLTK}: The natural language toolkit.
\newblock In \emph{Proceedings of the {ACL} Interactive Poster and Demonstration Sessions}, pages 214--217, Barcelona, Spain, 2004. Association for Computational Linguistics.

\bibitem[Busso et~al.(2008)Busso, Bulut, Lee, Kazemzadeh, Mower, Kim, Chang, Lee, and Narayanan]{busso2008iemocap}
Carlos Busso, Murtaza Bulut, Chi-Chun Lee, Abe Kazemzadeh, Emily Mower, Samuel Kim, Jeannette~N Chang, Sungbok Lee, and Shrikanth~S Narayanan.
\newblock Iemocap: Interactive emotional dyadic motion capture database.
\newblock \emph{Language resources and evaluation}, 42\penalty0 (4):\penalty0 335--359, 2008.

\bibitem[Cao et~al.(2014)Cao, Cooper, Keutmann, Gur, Nenkova, and Verma]{cao2014crema}
Houwei Cao, David~G Cooper, Michael~K Keutmann, Ruben~C Gur, Ani Nenkova, and Ragini Verma.
\newblock Crema-d: Crowd-sourced emotional multimodal actors dataset.
\newblock \emph{IEEE transactions on affective computing}, 5\penalty0 (4):\penalty0 377--390, 2014.

\bibitem[Chen et~al.(2020)Chen, Xie, Vedaldi, and Zisserman]{vggsound}
Honglie Chen, Weidi Xie, Andrea Vedaldi, and Andrew Zisserman.
\newblock Vggsound: A large-scale audio-visual dataset.
\newblock In \emph{ICASSP 2020-2020 IEEE International Conference on Acoustics, Speech and Signal Processing (ICASSP)}, pages 721--725. IEEE, 2020.

\bibitem[Chen et~al.(2022{\natexlab{a}})Chen, Du, Zhu, Ma, Berg-Kirkpatrick, and Dubnov]{HTSAT}
Ke Chen, Xingjian Du, Bilei Zhu, Zejun Ma, Taylor Berg-Kirkpatrick, and Shlomo Dubnov.
\newblock Hts-at: A hierarchical token-semantic audio transformer for sound classification and detection.
\newblock In \emph{ICASSP 2022-2022 IEEE International Conference on Acoustics, Speech and Signal Processing (ICASSP)}, pages 646--650. IEEE, 2022{\natexlab{a}}.

\bibitem[Chen et~al.(2022{\natexlab{b}})Chen, Wang, Chen, Wu, Liu, Chen, Li, Kanda, Yoshioka, Xiao, et~al.]{wavlm}
Sanyuan Chen, Chengyi Wang, Zhengyang Chen, Yu Wu, Shujie Liu, Zhuo Chen, Jinyu Li, Naoyuki Kanda, Takuya Yoshioka, Xiong Xiao, et~al.
\newblock Wavlm: Large-scale self-supervised pre-training for full stack speech processing.
\newblock \emph{IEEE Journal of Selected Topics in Signal Processing}, 16\penalty0 (6):\penalty0 1505--1518, 2022{\natexlab{b}}.

\bibitem[Chen et~al.(2023{\natexlab{a}})Chen, Wu, Wang, Liu, Tompkins, Chen, Che, Yu, and Wei]{BEATs}
Sanyuan Chen, Yu Wu, Chengyi Wang, Shujie Liu, Daniel Tompkins, Zhuo Chen, Wanxiang Che, Xiangzhan Yu, and Furu Wei.
\newblock Beats: Audio pre-training with acoustic tokenizers.
\newblock In \emph{International Conference on Machine Learning}, pages 5178--5193. PMLR, 2023{\natexlab{a}}.

\bibitem[Chen et~al.(2025)Chen, Xie, Chen, Zhao, Lee, Su, Sun, and Wang]{fusionaudio}
Shunian Chen, Xinyuan Xie, Zheshu Chen, Liyan Zhao, Owen Lee, Zhan Su, Qilin Sun, and Benyou Wang.
\newblock Fusionaudio-1.2 m: Towards fine-grained audio captioning with multimodal contextual fusion.
\newblock \emph{arXiv preprint arXiv:2506.01111}, 2025.

\bibitem[Chen et~al.(2023{\natexlab{b}})Chen, Wang, Lin, Qi, Ma, and Shan]{10.1609/aaai.v37i1.25113}
Yizhen Chen, Jie Wang, Lijian Lin, Zhongang Qi, Jin Ma, and Ying Shan.
\newblock Tagging before alignment: integrating multi-modal tags for video-text retrieval.
\newblock AAAI Press, 2023{\natexlab{b}}.

\bibitem[Chung et~al.(2018)Chung, Nagrani, and Zisserman]{voxceleb2}
Joon~Son Chung, Arsha Nagrani, and Andrew Zisserman.
\newblock Voxceleb2: Deep speaker recognition.
\newblock \emph{arXiv preprint arXiv:1806.05622}, 2018.

\bibitem[Cramer et~al.(2019)Cramer, Wu, Salamon, and Bello]{OpenL3}
Aurora~Linh Cramer, Ho-Hsiang Wu, Justin Salamon, and Juan~Pablo Bello.
\newblock Look, listen, and learn more: Design choices for deep audio embeddings.
\newblock In \emph{ICASSP 2019-2019 IEEE International Conference on Acoustics, Speech and Signal Processing (ICASSP)}, pages 3852--3856. IEEE, 2019.

\bibitem[Deng et~al.(2009)Deng, Dong, Socher, Li, Li, and Fei-Fei]{imagenet}
Jia Deng, Wei Dong, Richard Socher, Li-Jia Li, Kai Li, and Li Fei-Fei.
\newblock Imagenet: A large-scale hierarchical image database.
\newblock In \emph{2009 IEEE Conference on Computer Vision and Pattern Recognition}, pages 248--255, 2009.

\bibitem[Desplanques et~al.(2020)Desplanques, Thienpondt, and Demuynck]{ecapa}
Brecht Desplanques, Jenthe Thienpondt, and Kris Demuynck.
\newblock Ecapa-tdnn: Emphasized channel attention, propagation and aggregation in tdnn based speaker verification.
\newblock \emph{arXiv preprint arXiv:2005.07143}, 2020.

\bibitem[Devlin et~al.(2019)Devlin, Chang, Lee, and Toutanova]{devlin-etal-2019-bert}
Jacob Devlin, Ming-Wei Chang, Kenton Lee, and Kristina Toutanova.
\newblock {BERT}: Pre-training of deep bidirectional transformers for language understanding.
\newblock In \emph{Proceedings of the 2019 Conference of the North {A}merican Chapter of the Association for Computational Linguistics: Human Language Technologies, Volume 1 (Long and Short Papers)}, pages 4171--4186, Minneapolis, Minnesota, 2019. Association for Computational Linguistics.

\bibitem[Dinkel et~al.(2024)Dinkel, Yan, Wang, Zhang, Wang, and Wang]{Dasheng}
Heinrich Dinkel, Zhiyong Yan, Yongqing Wang, Junbo Zhang, Yujun Wang, and Bin Wang.
\newblock Scaling up masked audio encoder learning for general audio classification.
\newblock In \emph{Proc. Interspeech 2024}, pages 547--551, 2024.

\bibitem[Diwan et~al.(2025)Diwan, Zheng, Harwath, and Choi]{paraspeechcaps}
Anuj Diwan, Zhisheng Zheng, David Harwath, and Eunsol Choi.
\newblock Scaling rich style-prompted text-to-speech datasets.
\newblock \emph{arXiv preprint arXiv:2503.04713}, 2025.

\bibitem[Drossos et~al.(2020)Drossos, Lipping, and Virtanen]{clotho}
Konstantinos Drossos, Samuel Lipping, and Tuomas Virtanen.
\newblock Clotho: an audio captioning dataset.
\newblock In \emph{ICASSP 2020 - 2020 IEEE International Conference on Acoustics, Speech and Signal Processing (ICASSP)}, pages 736--740, 2020.

\bibitem[Elizalde et~al.(2023{\natexlab{a}})Elizalde, Deshmukh, Al~Ismail, and Wang]{CLAP2022}
Benjamin Elizalde, Soham Deshmukh, Mahmoud Al~Ismail, and Huaming Wang.
\newblock Clap learning audio concepts from natural language supervision.
\newblock In \emph{ICASSP 2023-2023 IEEE International Conference on Acoustics, Speech and Signal Processing (ICASSP)}, pages 1--5. IEEE, 2023{\natexlab{a}}.

\bibitem[Elizalde et~al.(2023{\natexlab{b}})Elizalde, Deshmukh, and Wang]{CLAP2023}
Benjamin Elizalde, Soham Deshmukh, and Huaming Wang.
\newblock Natural language supervision for general-purpose audio representations, 2023{\natexlab{b}}.

\bibitem[Engel et~al.(2017)Engel, Resnick, Roberts, Dieleman, Norouzi, Eck, and Simonyan]{NSynth}
Jesse Engel, Cinjon Resnick, Adam Roberts, Sander Dieleman, Mohammad Norouzi, Douglas Eck, and Karen Simonyan.
\newblock Neural audio synthesis of musical notes with wavenet autoencoders.
\newblock In \emph{International conference on machine learning}, pages 1068--1077. PMLR, 2017.

\bibitem[Fan et~al.(2023)Fan, Krishnan, Isola, Katabi, and Tian]{clip-rewrite}
Lijie Fan, Dilip Krishnan, Phillip Isola, Dina Katabi, and Yonglong Tian.
\newblock Improving clip training with language rewrites.
\newblock Red Hook, NY, USA, 2023. Curran Associates Inc.

\bibitem[Fonseca et~al.(2021)Fonseca, Favory, Pons, Font, and Serra]{fsd50k}
Eduardo Fonseca, Xavier Favory, Jordi Pons, Frederic Font, and Xavier Serra.
\newblock Fsd50k: an open dataset of human-labeled sound events.
\newblock \emph{IEEE/ACM Transactions on Audio, Speech, and Language Processing}, 30:\penalty0 829--852, 2021.

\bibitem[Gadre et~al.(2023)Gadre, Ilharco, Fang, Hayase, Smyrnis, Nguyen, Marten, Wortsman, Ghosh, Zhang, et~al.]{datacomp}
Samir~Yitzhak Gadre, Gabriel Ilharco, Alex Fang, Jonathan Hayase, Georgios Smyrnis, Thao Nguyen, Ryan Marten, Mitchell Wortsman, Dhruba Ghosh, Jieyu Zhang, et~al.
\newblock Datacomp: In search of the next generation of multimodal datasets.
\newblock \emph{Advances in Neural Information Processing Systems}, 36:\penalty0 27092--27112, 2023.

\bibitem[Gemmeke et~al.(2017)Gemmeke, Ellis, Freedman, Jansen, Lawrence, Moore, Plakal, and Ritter]{audioset}
Jort~F. Gemmeke, Daniel P.~W. Ellis, Dylan Freedman, Aren Jansen, Wade Lawrence, R.~Channing Moore, Manoj Plakal, and Marvin Ritter.
\newblock Audio set: An ontology and human-labeled dataset for audio events.
\newblock In \emph{2017 IEEE International Conference on Acoustics, Speech and Signal Processing (ICASSP)}, pages 776--780, 2017.

\bibitem[Goel et~al.(2025)Goel, Ghosh, Kim, Kumar, Kong, Lee, Yang, Duraiswami, Manocha, Valle, and Catanzaro]{af3}
Arushi Goel, Sreyan Ghosh, Jaehyeon Kim, Sonal Kumar, Zhifeng Kong, Sang-gil Lee, Chao-Han~Huck Yang, Ramani Duraiswami, Dinesh Manocha, Rafael Valle, and Bryan Catanzaro.
\newblock Audio flamingo 3: Advancing audio intelligence with fully open large audio language models.
\newblock \emph{arXiv preprint arXiv:2507.08128}, 2025.

\bibitem[Gong et~al.(2021)Gong, Chung, and Glass]{AST}
Yuan Gong, Yu-An Chung, and James Glass.
\newblock Ast: Audio spectrogram transformer.
\newblock In \emph{Proc. Interspeech 2021}, pages 571--575, 2021.

\bibitem[Gong et~al.(2022)Gong, Lai, Chung, and Glass]{ssast}
Yuan Gong, Cheng-I Lai, Yu-An Chung, and James Glass.
\newblock Ssast: Self-supervised audio spectrogram transformer.
\newblock In \emph{Proceedings of the AAAI Conference on Artificial Intelligence}, pages 10699--10709, 2022.

\bibitem[Gu et~al.(2018)Gu, Bradbury, Xiong, Li, and Socher]{gu2018nonautoregressive}
Jiatao Gu, James Bradbury, Caiming Xiong, Victor~O.K. Li, and Richard Socher.
\newblock Non-autoregressive neural machine translation.
\newblock In \emph{International Conference on Learning Representations}, 2018.

\bibitem[Guzhov et~al.(2021)Guzhov, Raue, Hees, and Dengel]{guzhov2021audioclip}
Andrey Guzhov, Federico Raue, Jörn Hees, and Andreas Dengel.
\newblock Audioclip: Extending clip to image, text and audio, 2021.

\bibitem[He et~al.(2016)He, Zhang, Ren, and Sun]{residual}
Kaiming He, Xiangyu Zhang, Shaoqing Ren, and Jian Sun.
\newblock Deep residual learning for image recognition.
\newblock In \emph{2016 IEEE Conference on Computer Vision and Pattern Recognition (CVPR)}, pages 770--778, 2016.

\bibitem[Hershey et~al.(2017)Hershey, Chaudhuri, Ellis, Gemmeke, Jansen, Moore, Plakal, Platt, Saurous, Seybold, et~al.]{VGGish}
Shawn Hershey, Sourish Chaudhuri, Daniel~PW Ellis, Jort~F Gemmeke, Aren Jansen, R~Channing Moore, Manoj Plakal, Devin Platt, Rif~A Saurous, Bryan Seybold, et~al.
\newblock Cnn architectures for large-scale audio classification.
\newblock In \emph{2017 ieee international conference on acoustics, speech and signal processing (icassp)}, pages 131--135. IEEE, 2017.

\bibitem[Hershey et~al.(2021)Hershey, Ellis, Fonseca, Jansen, Liu, Moore, and Plakal]{audiosetstrong}
Shawn Hershey, Daniel~PW Ellis, Eduardo Fonseca, Aren Jansen, Caroline Liu, R~Channing Moore, and Manoj Plakal.
\newblock The benefit of temporally-strong labels in audio event classification.
\newblock In \emph{ICASSP 2021-2021 IEEE International Conference on Acoustics, Speech and Signal Processing (ICASSP)}, pages 366--370. IEEE, 2021.

\bibitem[{Horizon Team, MiLM Plus}(2025)]{midashenglm7b}
{Horizon Team, MiLM Plus}.
\newblock Midashenglm: Efficient audio understanding with general audio captions.
\newblock Technical report, Xiaomi Inc., 2025.
\newblock Contributors: Heinrich Dinkel et al. (listed alphabetically in Appendix B).

\bibitem[Hsu et~al.(2021)Hsu, Bolte, Tsai, Lakhotia, Salakhutdinov, and Mohamed]{hsu2021hubert}
Wei-Ning Hsu, Benjamin Bolte, Yao-Hung~Hubert Tsai, Kushal Lakhotia, Ruslan Salakhutdinov, and Abdelrahman Mohamed.
\newblock Hubert: Self-supervised speech representation learning by masked prediction of hidden units.
\newblock \emph{IEEE/ACM transactions on audio, speech, and language processing}, 29:\penalty0 3451--3460, 2021.

\bibitem[Huang et~al.(2022)Huang, Xu, Li, Baevski, Auli, Galuba, Metze, and Feichtenhofer]{audioMAE}
Po-Yao Huang, Hu Xu, Juncheng Li, Alexei Baevski, Michael Auli, Wojciech Galuba, Florian Metze, and Christoph Feichtenhofer.
\newblock Masked autoencoders that listen.
\newblock \emph{Advances in Neural Information Processing Systems}, 35:\penalty0 28708--28720, 2022.

\bibitem[Huang et~al.(2024{\natexlab{a}})Huang, Zhang, Ma, Tian, Feng, Zhang, Li, Guo, and Zhang]{huang2024tag2text}
Xinyu Huang, Youcai Zhang, Jinyu Ma, Weiwei Tian, Rui Feng, Yuejie Zhang, Yaqian Li, Yandong Guo, and Lei Zhang.
\newblock Tag2text: Guiding vision-language model via image tagging.
\newblock In \emph{The Twelfth International Conference on Learning Representations}, 2024{\natexlab{a}}.

\bibitem[Huang et~al.(2024{\natexlab{b}})Huang, Ye, Kang, Feng, and Fan]{superclass_huang}
Zilong Huang, Qinghao Ye, Bingyi Kang, Jiashi Feng, and Haoqi Fan.
\newblock Classification done right for vision-language pre-training.
\newblock In \emph{NeurIPS}, 2024{\natexlab{b}}.

\bibitem[Jia et~al.(2021)Jia, Yang, Xia, Chen, Parekh, Pham, Le, Sung, Li, and Duerig]{pmlr-v139-jia21b}
Chao Jia, Yinfei Yang, Ye Xia, Yi-Ting Chen, Zarana Parekh, Hieu Pham, Quoc Le, Yun-Hsuan Sung, Zhen Li, and Tom Duerig.
\newblock Scaling up visual and vision-language representation learning with noisy text supervision.
\newblock In \emph{Proceedings of the 38th International Conference on Machine Learning}, pages 4904--4916. PMLR, 2021.

\bibitem[Kim et~al.(2019{\natexlab{a}})Kim, Kim, Lee, and Kim]{audiocaps}
Chris~Dongjoo Kim, Byeongchang Kim, Hyunmin Lee, and Gunhee Kim.
\newblock Audiocaps: Generating captions for audios in the wild.
\newblock In \emph{Proceedings of the 2019 Conference of the North American Chapter of the Association for Computational Linguistics: Human Language Technologies, Volume 1 (Long and Short Papers)}, pages 119--132, 2019{\natexlab{a}}.

\bibitem[Kim et~al.(2019{\natexlab{b}})Kim, Kim, Lee, and Kim]{kim-NAACL-HLT-2019}
Chris~Dongjoo Kim, Byeongchang Kim, Hyunmin Lee, and Gunhee Kim.
\newblock Audiocaps: Generating captions for audios in the wild.
\newblock In \emph{NAACL-HLT}, 2019{\natexlab{b}}.

\bibitem[Kong et~al.(2020)Kong, Cao, Iqbal, Wang, Wang, and Plumbley]{Panns}
Qiuqiang Kong, Yin Cao, Turab Iqbal, Yuxuan Wang, Wenwu Wang, and Mark~D. Plumbley.
\newblock Panns: Large-scale pretrained audio neural networks for audio pattern recognition.
\newblock \emph{IEEE/ACM Trans. Audio, Speech and Lang. Proc.}, 28:\penalty0 2880–2894, 2020.

\bibitem[Krizhevsky et~al.(2012)Krizhevsky, Sutskever, and Hinton]{imagenet-cls}
Alex Krizhevsky, Ilya Sutskever, and Geoffrey~E Hinton.
\newblock Imagenet classification with deep convolutional neural networks.
\newblock In \emph{Advances in Neural Information Processing Systems}. Curran Associates, Inc., 2012.

\bibitem[Lewis et~al.(2020)Lewis, Liu, Goyal, Ghazvininejad, Mohamed, Levy, Stoyanov, and Zettlemoyer]{lewis-etal-2020-bart}
Mike Lewis, Yinhan Liu, Naman Goyal, Marjan Ghazvininejad, Abdelrahman Mohamed, Omer Levy, Veselin Stoyanov, and Luke Zettlemoyer.
\newblock {BART}: Denoising sequence-to-sequence pre-training for natural language generation, translation, and comprehension.
\newblock In \emph{Proceedings of the 58th Annual Meeting of the Association for Computational Linguistics}, pages 7871--7880, Online, 2020. Association for Computational Linguistics.

\bibitem[Li et~al.(2022)Li, Li, Xiong, and Hoi]{li2022blip}
Junnan Li, Dongxu Li, Caiming Xiong, and Steven Hoi.
\newblock Blip: Bootstrapping language-image pre-training for unified vision-language understanding and generation.
\newblock In \emph{ICML}, 2022.

\bibitem[Li and Li(2022)]{li2022atst}
Xian Li and Xiaofei Li.
\newblock Atst: Audio representation learning with teacher-student transformer.
\newblock \emph{arXiv preprint arXiv:2204.12076}, 2022.

\bibitem[Li et~al.(2024)Li, Yuan, Zhang, Ma, Chen, Yin, Xiao, Lin, Ragni, Benetos, et~al.]{MERT}
Yizhi Li, Ruibin Yuan, Ge Zhang, Yinghao Ma, Xingran Chen, Hanzhi Yin, Chenghao Xiao, Chenghua Lin, Anton Ragni, Emmanouil Benetos, et~al.
\newblock Mert: Acoustic music understanding model with large-scale self-supervised training.
\newblock In \emph{ICLR}, 2024.

\bibitem[Liu et~al.(2023)Liu, Zheng, Wei, Tong, Liu, Chen, Wang, and Shen]{liu2023tagalign}
Qinying Liu, Kecheng Zheng, Wu Wei, Zhan Tong, Yu Liu, Wei Chen, Zilei Wang, and Yujun Shen.
\newblock Tagalign: Improving vision-language alignment with multi-tag classification.
\newblock \emph{arXiv preprint arXiv:2312.14149}, 2023.

\bibitem[Liu et~al.(2019)Liu, Ott, Goyal, Du, Joshi, Chen, Levy, Lewis, Zettlemoyer, and Stoyanov]{roberta}
Yinhan Liu, Myle Ott, Naman Goyal, Jingfei Du, Mandar Joshi, Danqi Chen, Omer Levy, Mike Lewis, Luke Zettlemoyer, and Veselin Stoyanov.
\newblock Roberta: A robustly optimized bert pretraining approach, 2019.

\bibitem[Livingstone and Russo(2018)]{ravdess}
Steven~R Livingstone and Frank~A Russo.
\newblock The ryerson audio-visual database of emotional speech and song (ravdess): A dynamic, multimodal set of facial and vocal expressions in north american english.
\newblock \emph{PloS one}, 13\penalty0 (5):\penalty0 e0196391, 2018.

\bibitem[Maaten and Hinton(2008)]{tsne}
Laurens van~der Maaten and Geoffrey Hinton.
\newblock Visualizing data using t-sne.
\newblock \emph{Journal of machine learning research}, 9\penalty0 (Nov):\penalty0 2579--2605, 2008.

\bibitem[Mahajan et~al.(2018)Mahajan, Girshick, Ramanathan, He, Paluri, Li, Bharambe, and van~der Maaten]{Mahajan_2018_ECCV}
Dhruv Mahajan, Ross Girshick, Vignesh Ramanathan, Kaiming He, Manohar Paluri, Yixuan Li, Ashwin Bharambe, and Laurens van~der Maaten.
\newblock Exploring the limits of weakly supervised pretraining.
\newblock In \emph{Proceedings of the European Conference on Computer Vision (ECCV)}, 2018.

\bibitem[Mehta et~al.(2024)Mehta, Horton, Faghri, Sekhavat, Najibi, Farajtabar, Tuzel, and Rastegari]{mehta2024catlipcliplevelvisualrecognition}
Sachin Mehta, Maxwell Horton, Fartash Faghri, Mohammad~Hossein Sekhavat, Mahyar Najibi, Mehrdad Farajtabar, Oncel Tuzel, and Mohammad Rastegari.
\newblock Catlip: Clip-level visual recognition accuracy with 2.7x faster pre-training on web-scale image-text data, 2024.

\bibitem[Mei et~al.(2024{\natexlab{a}})Mei, Meng, Liu, Kong, Ko, Zhao, Plumbley, Zou, and Wang]{mei2023wavcaps}
Xinhao Mei, Chutong Meng, Haohe Liu, Qiuqiang Kong, Tom Ko, Chengqi Zhao, Mark~D. Plumbley, Yuexian Zou, and Wenwu Wang.
\newblock Wav{C}aps: A {ChatGPT}-assisted weakly-labelled audio captioning dataset for audio-language multimodal research.
\newblock \emph{IEEE/ACM Transactions on Audio, Speech, and Language Processing}, pages 1--15, 2024{\natexlab{a}}.

\bibitem[Mei et~al.(2024{\natexlab{b}})Mei, Meng, Liu, Kong, Ko, Zhao, Plumbley, Zou, and Wang]{wavcaps}
Xinhao Mei, Chutong Meng, Haohe Liu, Qiuqiang Kong, Tom Ko, Chengqi Zhao, Mark~D Plumbley, Yuexian Zou, and Wenwu Wang.
\newblock Wavcaps: A chatgpt-assisted weakly-labelled audio captioning dataset for audio-language multimodal research.
\newblock \emph{IEEE/ACM Transactions on Audio, Speech, and Language Processing}, 32:\penalty0 3339--3354, 2024{\natexlab{b}}.

\bibitem[Niizumi et~al.(2024)Niizumi, Takeuchi, Ohishi, Harada, Yasuda, Tsubaki, and Imoto]{niizumi2024m2d}
Daisuke Niizumi, Daiki Takeuchi, Yasunori Ohishi, Noboru Harada, Masahiro Yasuda, Shunsuke Tsubaki, and Keisuke Imoto.
\newblock M2d-clap: Masked modeling duo meets clap for learning general-purpose audio-language representation.
\newblock \emph{arXiv preprint arXiv:2406.02032}, 2024.

\bibitem[Niizumi et~al.(2025)Niizumi, Takeuchi, Yasuda, Nguyen, Ohishi, and Harada]{niizumi2025m2d2}
Daisuke Niizumi, Daiki Takeuchi, Masahiro Yasuda, Binh~Thien Nguyen, Yasunori Ohishi, and Noboru Harada.
\newblock M2d2: Exploring general-purpose audio-language representations beyond clap.
\newblock \emph{arXiv preprint arXiv:2503.22104}, 2025.

\bibitem[OpenAI(2024)]{openai2024gpt4technicalreport}
OpenAI.
\newblock Gpt-4 technical report, 2024.

\bibitem[Poria et~al.(2018)Poria, Hazarika, Majumder, Naik, Cambria, and Mihalcea]{poria2018meld}
Soujanya Poria, Devamanyu Hazarika, Navonil Majumder, Gautam Naik, Erik Cambria, and Rada Mihalcea.
\newblock Meld: A multimodal multi-party dataset for emotion recognition in conversations.
\newblock \emph{arXiv preprint arXiv:1810.02508}, 2018.

\bibitem[Radford et~al.(2021)Radford, Kim, Hallacy, Ramesh, Goh, Agarwal, Sastry, Askell, Mishkin, Clark, et~al.]{CLIP}
Alec Radford, Jong~Wook Kim, Chris Hallacy, Aditya Ramesh, Gabriel Goh, Sandhini Agarwal, Girish Sastry, Amanda Askell, Pamela Mishkin, Jack Clark, et~al.
\newblock Learning transferable visual models from natural language supervision.
\newblock In \emph{International conference on machine learning}, pages 8748--8763. PmLR, 2021.

\bibitem[Radford et~al.(2023)Radford, Kim, Xu, Brockman, McLeavey, and Sutskever]{whisper}
Alec Radford, Jong~Wook Kim, Tao Xu, Greg Brockman, Christine McLeavey, and Ilya Sutskever.
\newblock Robust speech recognition via large-scale weak supervision.
\newblock In \emph{Proceedings of the 40th International Conference on Machine Learning}, pages 28492--28518, 2023.

\bibitem[Reimers and Gurevych(2019)]{sentencebert}
Nils Reimers and Iryna Gurevych.
\newblock Sentence-bert: Sentence embeddings using siamese bert-networks.
\newblock \emph{arXiv preprint arXiv:1908.10084}, 2019.

\bibitem[Rombach et~al.(2021)Rombach, Blattmann, Lorenz, Esser, and Ommer]{rombach2021highresolution}
Robin Rombach, Andreas Blattmann, Dominik Lorenz, Patrick Esser, and Björn Ommer.
\newblock High-resolution image synthesis with latent diffusion models, 2021.

\bibitem[Roy et~al.(2025)Roy, Liu, Lu, and Herremans]{jamendomaxcaps}
Abhinaba Roy, Renhang Liu, Tongyu Lu, and Dorien Herremans.
\newblock Jamendomaxcaps: A large scale music-caption dataset with imputed metadata.
\newblock \emph{arXiv preprint arXiv:2502.07461}, 2025.

\bibitem[Schuhmann et~al.(2022)Schuhmann, Beaumont, Vencu, Gordon, Wightman, Cherti, Coombes, Katta, Mullis, Wortsman, et~al.]{LAION5B}
Christoph Schuhmann, Romain Beaumont, Richard Vencu, Cade Gordon, Ross Wightman, Mehdi Cherti, Theo Coombes, Aarush Katta, Clayton Mullis, Mitchell Wortsman, et~al.
\newblock Laion-5b: An open large-scale dataset for training next generation image-text models.
\newblock \emph{Advances in neural information processing systems}, 35:\penalty0 25278--25294, 2022.

\bibitem[Sharma et~al.(2018)Sharma, Ding, Goodman, and Soricut]{conceptual-caption}
Piyush Sharma, Nan Ding, Sebastian Goodman, and Radu Soricut.
\newblock Conceptual captions: A cleaned, hypernymed, image alt-text dataset for automatic image captioning.
\newblock In \emph{ACL (1)}, pages 2556--2565. Association for Computational Linguistics, 2018.

\bibitem[Snyder et~al.(2018)Snyder, Garcia-Romero, Sell, Povey, and Khudanpur]{xvector}
David Snyder, Daniel Garcia-Romero, Gregory Sell, Daniel Povey, and Sanjeev Khudanpur.
\newblock X-vectors: Robust dnn embeddings for speaker recognition.
\newblock In \emph{2018 IEEE international conference on acoustics, speech and signal processing (ICASSP)}, pages 5329--5333. IEEE, 2018.

\bibitem[Sparck~Jones(1972)]{tf-idf}
Karen Sparck~Jones.
\newblock A statistical interpretation of term specificity and its application in retrieval.
\newblock \emph{Journal of documentation}, 28\penalty0 (1):\penalty0 11--21, 1972.

\bibitem[Su et~al.(2025)Su, Bai, Xu, Xu, and Dou]{ALPsurvey}
Yi Su, Jisheng Bai, Qisheng Xu, Kele Xu, and Yong Dou.
\newblock Audio-language models for audio-centric tasks: A survey.
\newblock \emph{arXiv preprint arXiv:2501.15177}, 2025.

\bibitem[Sun et~al.(2017)Sun, Shrivastava, Singh, and Gupta]{data-dl-era}
Chen Sun, Abhinav Shrivastava, Saurabh Singh, and Abhinav Gupta.
\newblock Revisiting unreasonable effectiveness of data in deep learning era.
\newblock In \emph{2017 IEEE International Conference on Computer Vision (ICCV)}, pages 843--852, 2017.

\bibitem[Sun et~al.(2024)Sun, Xu, Wu, and Xie]{autoacd}
Luoyi Sun, Xuenan Xu, Mengyue Wu, and Weidi Xie.
\newblock Auto-acd: A large-scale dataset for audio-language representation learning.
\newblock In \emph{Proceedings of the 32nd ACM International Conference on Multimedia}, pages 5025--5034, 2024.

\bibitem[Tschannen et~al.(2023)Tschannen, Kumar, Steiner, Zhai, Houlsby, and Beyer]{tschannen2023cappa}
Michael Tschannen, Manoj Kumar, Andreas~Peter Steiner, Xiaohua Zhai, Neil Houlsby, and Lucas Beyer.
\newblock Image captioners are scalable vision learners too.
\newblock In \emph{Thirty-seventh Conference on Neural Information Processing Systems}, 2023.

\bibitem[Tschannen et~al.(2025)Tschannen, Gritsenko, Wang, Naeem, Alabdulmohsin, Parthasarathy, Evans, Beyer, Xia, Mustafa, Hénaff, Harmsen, Steiner, and Zhai]{siglip2}
Michael Tschannen, Alexey Gritsenko, Xiao Wang, Muhammad~Ferjad Naeem, Ibrahim Alabdulmohsin, Nikhil Parthasarathy, Talfan Evans, Lucas Beyer, Ye Xia, Basil Mustafa, Olivier Hénaff, Jeremiah Harmsen, Andreas Steiner, and Xiaohua Zhai.
\newblock Siglip 2: Multilingual vision-language encoders with improved semantic understanding, localization, and dense features, 2025.

\bibitem[Tseng et~al.(2025)Tseng, Zhou, Huo, Shao, Zhang, and Yu]{anonymous2025revisiting}
Wei-Cheng Tseng, Xuanru Zhou, Mingyue Huo, Yiwen Shao, Hao Zhang, and Dong Yu.
\newblock Revisiting audio-language pretraining for learning general-purpose audio representation, 2025.

\bibitem[Turian et~al.(2022)Turian, Shier, Khan, Raj, Schuller, Steinmetz, Malloy, Tzanetakis, Velarde, McNally, Henry, Pinto, Noufi, Clough, Herremans, Fonseca, Engel, Salamon, Esling, Manocha, Watanabe, Jin, and Bisk]{hear}
Joseph Turian, Jordie Shier, Humair~Raj Khan, Bhiksha Raj, Björn~W. Schuller, Christian~J. Steinmetz, Colin Malloy, George Tzanetakis, Gissel Velarde, Kirk McNally, Max Henry, Nicolas Pinto, Camille Noufi, Christian Clough, Dorien Herremans, Eduardo Fonseca, Jesse Engel, Justin Salamon, Philippe Esling, Pranay Manocha, Shinji Watanabe, Zeyu Jin, and Yonatan Bisk.
\newblock Hear: Holistic evaluation of audio representations, 2022.

\bibitem[van~den Oord et~al.(2019)van~den Oord, Li, and Vinyals]{infonce}
Aaron van~den Oord, Yazhe Li, and Oriol Vinyals.
\newblock Representation learning with contrastive predictive coding, 2019.

\bibitem[wen Yang et~al.(2021)wen Yang, Chi, Chuang, Lai, Lakhotia, Lin, Liu, Shi, Chang, Lin, Huang, Tseng, tik Lee, Liu, Huang, Dong, Li, Watanabe, Mohamed, and yi~Lee]{superb}
Shu wen Yang, Po-Han Chi, Yung-Sung Chuang, Cheng-I~Jeff Lai, Kushal Lakhotia, Yist~Y. Lin, Andy~T. Liu, Jiatong Shi, Xuankai Chang, Guan-Ting Lin, Tzu-Hsien Huang, Wei-Cheng Tseng, Ko tik Lee, Da-Rong Liu, Zili Huang, Shuyan Dong, Shang-Wen Li, Shinji Watanabe, Abdelrahman Mohamed, and Hung yi Lee.
\newblock Superb: Speech processing universal performance benchmark.
\newblock In \emph{Interspeech 2021}, pages 1194--1198, 2021.

\bibitem[Wolff and Weyde(2012)]{MagnaTagATune}
Daniel Wolff and Tillman Weyde.
\newblock Adapting similarity on the magnatagatune database: effects of model and feature choices.
\newblock page 931–936, New York, NY, USA, 2012. Association for Computing Machinery.

\bibitem[Wu et~al.(2022)Wu, Seetharaman, Kumar, and Bello]{learn-from-clip}
Ho-Hsiang Wu, Prem Seetharaman, Kundan Kumar, and Juan~Pablo Bello.
\newblock Wav2clip: Learning robust audio representations from clip.
\newblock In \emph{ICASSP 2022 - 2022 IEEE International Conference on Acoustics, Speech and Signal Processing (ICASSP)}, pages 4563--4567, 2022.

\bibitem[Wu* et~al.(2023)Wu*, Chen*, Zhang*, Hui*, Berg-Kirkpatrick, and Dubnov]{laionclap2023}
Yusong Wu*, Ke Chen*, Tianyu Zhang*, Yuchen Hui*, Taylor Berg-Kirkpatrick, and Shlomo Dubnov.
\newblock Large-scale contrastive language-audio pretraining with feature fusion and keyword-to-caption augmentation.
\newblock In \emph{IEEE International Conference on Acoustics, Speech and Signal Processing, ICASSP}, 2023.

\bibitem[Xu et~al.(2022)Xu, De~Mello, Liu, Byeon, Breuel, Kautz, and Wang]{Xu_2022_CVPR}
Jiarui Xu, Shalini De~Mello, Sifei Liu, Wonmin Byeon, Thomas Breuel, Jan Kautz, and Xiaolong Wang.
\newblock Groupvit: Semantic segmentation emerges from text supervision.
\newblock In \emph{Proceedings of the IEEE/CVF Conference on Computer Vision and Pattern Recognition (CVPR)}, pages 18134--18144, 2022.

\bibitem[Xu et~al.(2023{\natexlab{a}})Xu, Hou, Zhang, Feng, Wang, Qiao, and Xie]{xu2023learning}
Jilan Xu, Junlin Hou, Yuejie Zhang, Rui Feng, Yi Wang, Yu Qiao, and Weidi Xie.
\newblock Learning open-vocabulary semantic segmentation models from natural language supervision.
\newblock In \emph{Proceedings of the IEEE/CVF Conference on Computer Vision and Pattern Recognition}, pages 2935--2944, 2023{\natexlab{a}}.

\bibitem[Xu et~al.(2025)Xu, Guo, Hu, Chu, Wang, He, Wang, Shi, He, Zhu, Lv, Wang, Guo, Wang, Ma, Zhang, Zhang, Hao, Guo, Yang, Zhang, Ma, Wei, Bai, Chen, Liu, Wang, Yang, Liu, Ren, Zheng, Men, Zhou, Yu, Yang, Yu, Zhou, and Lin]{Qwen3-Omni}
Jin Xu, Zhifang Guo, Hangrui Hu, Yunfei Chu, Xiong Wang, Jinzheng He, Yuxuan Wang, Xian Shi, Ting He, Xinfa Zhu, Yuanjun Lv, Yongqi Wang, Dake Guo, He Wang, Linhan Ma, Pei Zhang, Xinyu Zhang, Hongkun Hao, Zishan Guo, Baosong Yang, Bin Zhang, Ziyang Ma, Xipin Wei, Shuai Bai, Keqin Chen, Xuejing Liu, Peng Wang, Mingkun Yang, Dayiheng Liu, Xingzhang Ren, Bo Zheng, Rui Men, Fan Zhou, Bowen Yu, Jianxin Yang, Le Yu, Jingren Zhou, and Junyang Lin.
\newblock Qwen3-omni technical report.
\newblock \emph{arXiv preprint arXiv:2509.17765}, 2025.

\bibitem[Xu et~al.(2015)Xu, Ba, Kiros, Cho, Courville, Salakhutdinov, Zemel, and Bengio]{show-and-tell}
Kelvin Xu, Jimmy~Lei Ba, Ryan Kiros, Kyunghyun Cho, Aaron Courville, Ruslan Salakhutdinov, Richard~S. Zemel, and Yoshua Bengio.
\newblock Show, attend and tell: neural image caption generation with visual attention.
\newblock In \emph{Proceedings of the 32nd International Conference on International Conference on Machine Learning - Volume 37}, page 2048–2057. JMLR.org, 2015.

\bibitem[Xu et~al.(2023{\natexlab{b}})Xu, Zhang, Zhou, Zhang, Xie, Wu, and Zhu]{xu2023blat}
Xuenan Xu, Zhiling Zhang, Zelin Zhou, Pingyue Zhang, Zeyu Xie, Mengyue Wu, and Kenny~Q Zhu.
\newblock Blat: Bootstrapping language-audio pre-training based on audioset tag-guided synthetic data.
\newblock In \emph{Proceedings of the 31st ACM International Conference on Multimedia}, pages 2756--2764, 2023{\natexlab{b}}.

\bibitem[Yang et~al.(2025)Yang, Yang, Zhang, Hui, Zheng, Yu, Li, Liu, Huang, Wei, Lin, Yang, Tu, Zhang, Yang, Yang, Zhou, Lin, Dang, Lu, Bao, Yang, Yu, Li, Xue, Zhang, Zhu, Men, Lin, Li, Tang, Xia, Ren, Ren, Fan, Su, Zhang, Wan, Liu, Cui, Zhang, and Qiu]{qwen2025qwen25technicalreport}
An Yang, Baosong Yang, Beichen Zhang, Binyuan Hui, Bo Zheng, Bowen Yu, Chengyuan Li, Dayiheng Liu, Fei Huang, Haoran Wei, Huan Lin, Jian Yang, Jianhong Tu, Jianwei Zhang, Jianxin Yang, Jiaxi Yang, Jingren Zhou, Junyang Lin, Kai Dang, Keming Lu, Keqin Bao, Kexin Yang, Le Yu, Mei Li, Mingfeng Xue, Pei Zhang, Qin Zhu, Rui Men, Runji Lin, Tianhao Li, Tianyi Tang, Tingyu Xia, Xingzhang Ren, Xuancheng Ren, Yang Fan, Yang Su, Yichang Zhang, Yu Wan, Yuqiong Liu, Zeyu Cui, Zhenru Zhang, and Zihan Qiu.
\newblock Qwen2.5 technical report, 2025.

\bibitem[Yang et~al.(2024)Yang, Xu, Liu, Chu, Jiang, Zhou, Leng, Lv, Zhao, Zhou, and Zhou]{airbench}
Qian Yang, Jin Xu, Wenrui Liu, Yunfei Chu, Ziyue Jiang, Xiaohuan Zhou, Yichong Leng, Yuanjun Lv, Zhou Zhao, Chang Zhou, and Jingren Zhou.
\newblock {AIR}-bench: Benchmarking large audio-language models via generative comprehension.
\newblock In \emph{Proceedings of the 62nd Annual Meeting of the Association for Computational Linguistics (Volume 1: Long Papers)}, pages 1979--1998, Bangkok, Thailand, 2024. Association for Computational Linguistics.

\bibitem[Yao et~al.(2024)Yao, Guo, Yang, Kang, Kuang, Yang, Jin, Lin, and Povey]{yao2024zipformer}
Zengwei Yao, Liyong Guo, Xiaoyu Yang, Wei Kang, Fangjun Kuang, Yifan Yang, Zengrui Jin, Long Lin, and Daniel Povey.
\newblock Zipformer: A faster and better encoder for automatic speech recognition.
\newblock In \emph{The Twelfth International Conference on Learning Representations}, 2024.

\bibitem[Yuan et~al.(2023)Yuan, Ma, Li, Zhang, Chen, Yin, Liu, Huang, Tian, Deng, et~al.]{marble}
Ruibin Yuan, Yinghao Ma, Yizhi Li, Ge Zhang, Xingran Chen, Hanzhi Yin, Yiqi Liu, Jiawen Huang, Zeyue Tian, Binyue Deng, et~al.
\newblock Marble: Music audio representation benchmark for universal evaluation.
\newblock \emph{Advances in Neural Information Processing Systems}, 36:\penalty0 39626--39647, 2023.

\bibitem[Zhai et~al.(2023)Zhai, Mustafa, Kolesnikov, and Beyer]{siglip}
Xiaohua Zhai, Basil Mustafa, Alexander Kolesnikov, and Lucas Beyer.
\newblock Sigmoid loss for language image pre-training.
\newblock In \emph{2023 IEEE/CVF International Conference on Computer Vision (ICCV)}, pages 11941--11952, 2023.

\bibitem[Zhang and Zhou(2014)]{mtc}
Min-Ling Zhang and Zhi-Hua Zhou.
\newblock A review on multi-label learning algorithms.
\newblock \emph{IEEE Transactions on Knowledge and Data Engineering}, 26\penalty0 (8):\penalty0 1819--1837, 2014.

\bibitem[Zhu et~al.(2024)Zhu, Darefsky, and Duan]{zhu2024cacophony}
Ge Zhu, Jordan Darefsky, and Zhiyao Duan.
\newblock Cacophony: An improved contrastive audio-text model.
\newblock \emph{IEEE/ACM Transactions on Audio, Speech, and Language Processing}, 2024.

\bibitem[Zhu et~al.(2025)Zhu, Zhou, Chen, Yu, Ma, Gu, Luo, Tan, and Chen]{muq}
Haina Zhu, Yizhi Zhou, Hangting Chen, Jianwei Yu, Ziyang Ma, Rongzhi Gu, Yi Luo, Wei Tan, and Xie Chen.
\newblock Muq: Self-supervised music representation learning with mel residual vector quantization.
\newblock \emph{arXiv preprint arXiv:2501.01108}, 2025.

\bibitem[Żelasko et~al.(2021)Żelasko, Povey, Trmal, and Khudanpur]{lhotse}
Piotr Żelasko, Daniel Povey, Jan~"Yenda" Trmal, and Sanjeev Khudanpur.
\newblock Lhotse: a speech data representation library for the modern deep learning ecosystem, 2021.

\end{thebibliography}
